\documentclass[11pt,a4paper]{article}
\pdfoutput=1 

\usepackage{jheppub}
\usepackage[T1]{fontenc}
\usepackage[latin9]{inputenc}
\usepackage{esint}
\usepackage[mathscr]{euscript}
\usepackage{amsmath,enumerate}

\def\spaa#1.#2.#3{\langle\mskip-1mu{#1}|#2|{#3}\mskip-1mu\rangle}
\def\spbb#1.#2.#3{[\mskip-1mu{#1}|#2|{#3}\mskip-1mu]}
\def\spa#1.#2{\left\langle#1\,#2\right\rangle}
\def\spb#1.#2{\left[#1\,#2\right]}
\def\spab#1.#2.#3{\left\langle#1|#2|#3\right]}
\def\spba#1.#2.#3{\left[#1|#2|#3\right\rangle}

\def\P3b{\bar{P}_3}

\def\g0{\gamma_0}

\newcommand{\beq}{\begin{equation}}
\newcommand{\eeq}{\end{equation}}
\newcommand{\beqn}{\begin{eqnarray}}
\newcommand{\eeqn}{\end{eqnarray}}

\def\vec#1{{\mbox{\boldmath$#1$}}}

\newcommand{\as}{\alpha_s}

\newcommand\POWHEGBOX{{\tt POWHEG BOX}}
\newcommand\POWHEG{{\tt POWHEG}}
\newcommand\SHERPA{{\tt Sherpa}}
\newcommand\MCatNLO{{\tt MC@NLO}}
\newcommand\PYTHIA{{\tt PYTHIA}}

\newcommand\MINLO{MiNLO}

\newcommand\gghmass{{\tt gg\_H\_quark-mass-effects}}

\newcommand\MiNLO{{\tt MiNLO}}
\newcommand\HNNLO{{\tt HNNLO}}
\newcommand\NNLOPS{{\tt NNLOPS}}

\newcommand\HJ{{\tt HJ}}

\newcommand\mH{m_{\scriptscriptstyle\rm H}}
\newcommand\PhiB{\Phi_{\rm B}}
\newcommand\PhiR{\Phi_{\rm rad}}
\newcommand\mathd{{\mathrm d}}
\newcommand\MSBAR{\overline{\rm MS}}
\newcommand\DR{DR}

\preprint{\\\\MCnet-15-01\\CERN-PH-TH/2015-006}

\title{Finite quark-mass effects in the NNLOPS POWHEG+MiNLO Higgs generator}

\author[a]{Keith Hamilton,}
\author[b]{Paolo Nason,}
\author[c]{Giulia Zanderighi$^1$\note{On leave from Rudolf Peierls Centre for Theoretical Physics, University of Oxford, 1 Keble Road, UK}}

\affiliation[a]{Department of Physics and Astronomy, University College London,\\London, WC1E 6BT, UK}
\affiliation[b]{INFN, Sezione di Milano Bicocca, Piazza della Scienza 3, 20126 Milan, Italy}
\affiliation[c]{Theory Division, CERN, CH--1211, Geneva 23, Switzerland}

\emailAdd{keith.hamilton@ucl.ac.uk}
\emailAdd{paolo.nason@mib.infn.it}
\emailAdd{giulia.zanderighi@cern.ch}

\abstract{We include finite top- and bottom-mass effects in the
  next-to-next-to-leading order parton shower (NNLOPS) event generator
  for inclusive Higgs boson production in gluon fusion based upon the
  \POWHEG{}+\MiNLO{} approach.  Since fixed-order results for
  quark-mass effects only reach NLO accuracy, we add them to the
  NNLOPS generator at that accuracy. We explore uncertainties related
  to the unknown all-order logarithmic structure of bottom-mass
  effects by comparing the assumption of full exponentiation to no
  exponentiation at all.  Phenomenological results showing the effects
  of finite quark-masses in the NNLOPS simulation are presented. These
  suggest that the aforementioned uncertainty is well contained within
  the envelope of plain renormalization and factorization scale
  uncertainties.}


\begin{document}
\maketitle
\flushbottom
\section{Introduction}

Since the 2012 LHC discovery of a new spin-zero particle
\cite{Aad:2012tfa,Chatrchyan:2012ufa}, all subsequent experimental
analysis (see
e.g.~\cite{CMS:bxa,Chatrchyan:2013vaa,Chatrchyan:2014nva,CMS:zwa,Khachatryan:2014jba,ATLAS:2013oma,ATLAS:2013qma,ATLAS:2013rma,ATLAS-CONF-2013-040})
has revealed that its properties are very much consistent with those
predicted for the Standard Model (SM) Higgs boson.  The discovery of
what appears to be the SM Higgs boson, without any new physics below
the TeV scale to stabilize its mass, poses profound and challenging
questions for the theoretical community.  By the same token, LHC
studies to quantify the level of agreement between SM Higgs boson
predictions and experimental measurements, by increasing the precision
in both, has further intensified.

Thus far, tests and measurements of Higgs boson properties, in almost
all cases, have focused on rather inclusive observables, such as
production cross sections and branching ratios.
However, even for fully inclusive quantities experimental analysis do
often classify events in categories depending on the number of
accompanying jets, and, at times, jet-vetos are used to enhance the
signal-to-background ratio.  In general these measurements require the
use of fully exclusive Monte Carlo generators, possibly in combination
with dedicated, theoretical calculations.

For gluon-fusion Higgs production, first differential measurements
have been published in the second half of 2014 by the ATLAS
collaboration \cite{Aad:2014lwa,Aad:2014tca}.  These measurements are
at present statistically limited, however, forthcoming LHC Run-II data
will lead to substantial improvements, and by the end of Run-II the
measurement uncertainties are estimated to be substantially reduced.
Full exploitation of these data, in particular the degree to which we
are able to resolve or exclude new physics in the Higgs sector,
requires that all theoretical tools, including Monte Carlo
simulations, be as precise as possible.

Next-to-leading order calculations matched to parton showers (NLOPS)
\cite{Frixione:2002ik,Nason:2004rx,Frixione:2007vw} are now the
standard for fully exclusive Monte Carlo predictions and are
indispensable tools for the LHC experimental collaborations.

Recently, NNLOPS generators for
Higgs~\cite{Hamilton:2013fea,Hoche:2014dla} and Drell
Yan~\cite{Hoeche:2014aia,Karlberg:2014qua} production have also
appeared. In particular, the generator of ref.\cite{Hamilton:2013fea},
based upon the \MiNLO{} method~\cite{Hamilton:2012rf}, is particularly
attractive since it achieves NNLO accuracy without recourse to an
unphysical partitioning of phase space.
This generator is based upon the large $m_t$ approximation of the
Higgs coupling to gluons. On the other hand, it is known that finite
quark-mass effects are important especially in the case of production
in association to energetic jets, since the radiation resolves the
internal structure of the gluon-Higgs effective coupling.

Finite quark-mass effects to the total cross section are known exactly
at NLO $\left(\mathcal{O}\left(\alpha_{{\scriptscriptstyle
    \mathrm{S}}}^{3}\right)\right)$ in QCD perturbation theory
\cite{Graudenz:1992pv,Spira:1995rr,Harlander:2005rq,Anastasiou:2006hc,Aglietti:2006tp,Bonciani:2007ex}.
Finite quark-mass effects on high transverse momentum 
Higgs production were first calculated at LO
$\left(\mathcal{O}\left(\alpha_{{\scriptscriptstyle
    \mathrm{S}}}^{3}\right)\right)$ in
refs.~\cite{Ellis:1987xu,Baur:1989cm} and, as an expansion in $1/m_t$,
at NLO for finite top-quark-mass effects by Harlander et
al. \cite{Harlander:2012hf}.  Some of the former finite quark-mass
corrections have been implemented in public fixed order computer codes
e.g. MCFM \cite{MCFM} FehiPro
\cite{Anastasiou:2005qj,Anastasiou:2009kn} and SusHi
\cite{Harlander:2012pb}.

Finite quark-mass effects have been included in NLOPS Monte Carlo
simulations of Higgs boson production in \POWHEG{}
\cite{Bagnaschi:2011tu} and \MCatNLO{} version 4.08 onwards
\cite{Frixione:2002ik}.  Recently the \SHERPA{} collaboration have
introduced a treatment of finite top-mass effects
\cite{Buschmann:2014sia} in the context of their multi-jet NLOPS
merging scheme MEPS@NLO.
Differences between the predictions of the \POWHEG{} program of
ref.~\cite{Bagnaschi:2011tu} and \MCatNLO{}, in particular regarding
their response to the inclusion of finite $b$-quark-mass effects, have
been a source of discussion and have stimulated further analytic work
on the treatment of bottom-mass effects in resummed calculations.

The earliest analytic resummation work to include mass effects was
that of Mantler and Wiesemann at LO+NLL accuracy
\cite{Mantler:2012bj}, work which was later extended to include also
MSSM effects \cite{Harlander:2014uea}.  An NNLL+NNLO resummation of
the transverse momentum spectrum of the boson in the large-$m_{t}$
limit was subsequently augmented with finite $t$- and $b$-quark-mass
effects at the NLL+NLO level by Grazzini and Sargsyan
\cite{Grazzini:2013mca}. Shortly following this work Banfi et
al. presented an extension of their NNLL+NNLO large-$m_{t}$
computation of the efficiency for Higgs boson production in the
presence of a jet veto in ref.~\cite{Banfi:2013eda}. All of these
analytic resummation computations took a different approach to the
inclusion of finite $b$-quark-mass effects, in particular, to their
handling of the enhanced
$\sim\frac{1}{p_{T}^{2}}\,\frac{m_{b}^{2}}{m_{H}^{2}}\log^{2}\frac{m_{b}^{2}}{p_{T}^{2}}$
terms in the region $m_{b}<p_{T}<m_{H}$.

In this work we address the extension of the NNLOPS event generator
for Higgs boson production in ref.~\cite{Hamilton:2013fea} to include
finite top- and bottom-quark-mass effects. In section two we present
our theoretical rationale and method for doing this, with each
subsection describing the implementation of finite quark-mass effects
in a different layer of the event generator. Sec.~2.1 concerns the
required modifications at the level of the underlying \HJ{} NLOPS
generator \cite{Campbell:2012am}.  Sec.~2.2 enters into the discussion
on the role of finite quark-mass effects and resummation, describing
their implementation (or not) in the \MINLO{} Sudakov form factor and
the theoretical justification for our approach. Sec.~2.3 describes how
the finite quark-mass effects are accounted for in the NNLO
predictions required for the reweighting stage.  In Sec.~3 we present
a selection of numerical results obtained with the new NNLOPS
generator using various options to explore theoretical
uncertainties. Conclusions are drawn in Sec.~4.  Some technical
details relevant to our new implementations are illustrated in the
Appendix.

\section{Method}

In ref.~\cite{Hamilton:2013fea} a generator that is NNLO accurate and
includes parton shower effects (NNLOPS accurate from now on) was
presented for inclusive Higgs production.  This generator is based on
the large $m_t$ effective theory, in which the Higgs boson emitted
from a top-loop is approximated by an effective tree-level coupling of
the Higgs to gluons and the interaction of the Higgs boson to lighter
quarks (including the bottom quark) is neglected.

Our goal here is to correct this NNLOPS Higgs generator in such a way
that at least the most important mass effects are included. The NNLOPS
generator relies upon three main components:
\begin{itemize}
\item the \POWHEG{} \HJ{} generator of ref.~\cite{Campbell:2012am};
\item the \MiNLO{} procedure on \HJ{} discussed in ref.~\cite{Hamilton:2012rf};
\item the NNLO fixed order calculation of ref.~\cite{Grazzini:2013mca};
\item the reweighting procedure described in ref.~\cite{Hamilton:2013fea}.
\end{itemize}
In the following we describe how, and to what extent, we include mass
effects in the first three items of this list. The reweighting
procedure remains obviously the same.

\subsection{Mass effects in the \HJ{} generator}\label{sec:meHJ}
An NLO calculation (i.e.~of order $\as^4$) of Higgs plus jet
production including the finite quark-mass effects is not available at
the moment, and we will thus resort to the following approximations.
We begin by considering the quark-mass corrections due to the top loop
only.  We multiply all components (Born, virtual and real) of the
infinite top mass approximation formulae by the ratio of the matrix
elements for the production of a Higgs boson in association with a
light parton, including exact one loop top mass dependence, divided by
the same matrix element in the infinite top mass approximation. These
matrix elements are all evaluated at the Born kinematics for the Born
and virtual contribution, and at the underlying Born kinematics for
the real one.\footnote{The underlying Born kinematics of a given real
  emission kinematics configuration is obtained by an $n+1$ to $n$
  mapping procedure (specified in details in
  ref.~\cite{Nason:2004rx,Frixione:2007vw}) that is such that in the
  singular collinear limits corresponds to the Born kinematics of the
  factorized cross section.}  This procedure guarantees that at the
Born level the cross section has the full top mass dependence. The
real emission cross section in the soft and collinear limits, has also
fully corrected top mass dependence, and the same holds for the part
of the virtual corrections that arises from virtual gluons of energy
much below the top mass.  Thus, the contributions that we miss have to
do either with real emissions with widely separated jets, or virtual
corrections where the gluon energy is not small with respect to the
top mass.  These corrections are in fact of two-loop level, and are
not yet available.

The inclusion of bottom mass effects is more delicate. In this case,
in fact, there is still factorization for light parton emissions at
scale below the bottom mass. However, this scale is now much smaller
than the Higgs mass, and such factorization may turn out not to be
useful in most of the range of light parton emission, or of light
parton virtualities in one-loop corrections. We thus consider two
options in this case. In the first one we ignore this problem, and
include the effect of the bottom quark in our rescaling factor. In the
second option, we apply the top mass correction to all components of
the cross section, but include the bottom mass effect only in the Born
case.

The matrix elements for $H+$jet production at order $\as^3$ and with
full quark-mass dependence have been computed in
ref.~\cite{Ellis:1987xu,Baur:1989cm}, and are implemented in the
\POWHEGBOX{} in the generator \gghmass{}~\cite{Bagnaschi:2011tu}. We
have used the code of ref.~\cite{Bagnaschi:2011tu} for our purposes.

We perform our reweighting at the level of the calculation of the
matrix elements. The following reweighting factors are computed
together with the Born term:
\begin{equation}
K_{tb}=\frac{B_{tb}(\PhiB,m_t,m_b)}{B_{\rm inf}},\quad
K_{t}=\frac{B_{t}(\PhiB,m_t)}{B_{\rm inf}},\quad
\end{equation}
where
\begin{equation}
B_{\rm inf} = B_{tb}(\PhiB,m_t,m_b)|_{m_t=\infty,m_b=0}.
\end{equation}
These reweighting factors are stored, and the Born matrix elements are
multiplied by $K_{tb}$. The real and virtual corrections are
multiplied by $K_{tb}$ in our first option, and by $K_{t}$ in our
second option.\footnote{ The Born matrix elements that are used to
  compute the real counterterms, the collinear remnants and the soft
  collinear terms are always rescaled in the same way as the real and
  virtual terms.}

We notice that with this procedure, the Born matrix element is
computed with the exact quark-mass dependence. At variance with other
methods, however, also the higher order terms are reweighted point by
point in the phase space, either using the same reweighting factor, or
a factor including only top mass effects. Our motivation to do so is
that this is correct at least for the Born term and for the terms
proportional to it.

\subsection{Mass effects in the \MiNLO{} procedure.}
\MiNLO{}~\cite{Hamilton:2012np} is in essence and extension of the
well-known CKKW procedure~\cite{Catani:2001cc} at the NLO level. One
first associates a most likely branching history to the kinematic
structure of the event.  The hardness of the branchings are then used
to set the factorization and renormalization scales. Furthermore,
appropriate Sudakov form factors are supplied to the process, in full
analogy with shower algorithms. Care has to be taken to subtract NLO
terms arising from the expansion of the Sudakov form factor in order
to maintain NLO accuracy of the cross section.  The Sudakov form
factors guarantee that the \MiNLO{} improved cross section can be
integrated down to vanishing momenta of the associated jets.
Furthermore, in ref.~\cite{Hamilton:2012rf}, it was shown that, in the
case of the production of a boson in association with a jet, by a
suitable refinement of the procedure, one could achieve NLO
(i.e. $\as^3$) accuracy for fully integrated quantities, i.e.  even
without requiring the presence of the jet.

The \MiNLO{} Sudakov form factor of ref.~\cite{Hamilton:2012rf} is
computed assuming that the main production vertex is pointlike, which
is the case in the infinite top-mass approximation. This is a
justified assumption if we only consider the top loop in the Higgs
production vertex. Soft gluons are in fact characterized by energies
below the Higgs mass. The flow of their momentum through the top loop
is thus expected to have a very limited effect, in view of the large
value of the top mass.  On the other hand, also the bottom contributes
to the Higgs coupling, with the interference of the bottom and top
loop affecting the total cross section by about -7\%{}. Furthermore,
it was noticed that the \POWHEG{} generator \gghmass{} yielded a
transverse momentum distribution for the Higgs differing substantially
with the analogous~\MCatNLO{} simulation, and with several analytic
treatments of the Higgs transverse momentum spectrum.  These
differences are easily traced back to the fact that in \POWHEG{} the
Sudakov form factor is computed including the effects of the bottom
finite mass. It is in fact given by the formula
\begin{equation}\label{eq:powhegsud}
\exp\left[-\int_{p_T'>p_T} \frac{R(\PhiB,\PhiR')}{B(\PhiB)} \;\mathd \PhiR'
\right],
\end{equation}
where $R$ is the real emission cross section and $B$ is the Born term.
On the other hand, in standard resummation formulae the real cross
section is replaced by its Altarelli-Parisi
approximation~\cite{Altarelli:1977zs}. This assumes that the
underlying Born cross section is not affected by virtualities of the
incoming partons smaller than the Higgs mass. 

This issue was dealt with in different ways in the literature.  In
ref.~\cite{Grazzini:2013mca} it was assumed that the resummation scale
in the case of bottom mediated Higgs production should have been set
to the mass of the bottom rather than to the mass of the Higgs, in
order to satisfy the basic resummation assumption that the soft gluons
should be softer than the quark-mass.  On the other hand, in
ref.~\cite{Banfi:2013eda}, it was shown by a detailed analytical
study, that at order $\as^3$ the double logarithmic structure of the
gluon emission in Higgs production with quark-mass effects remains the
same as in the infinite mass limit.  There are however differences at
the single logarithmic level.  The all order logarithmic structure,
accounting for the complex three-scales problem induced by the bottom
contribution, is at present not known even at the double logarithmic
level. It is thus clear that this fact will introduce an uncertainty
in our current prediction that must be modeled in some way.  In the
present work, we do the following. We consider two possible approaches
to the computation of the Higgs Sudakov form factor. In the first
approach, that we will label as MEMB (standing for Matrix Elements
$m_b$), we keep the same \MiNLO{} Sudakov that we use in the
traditional \HJ{}-\MiNLO{} generator, thus assuming that no large
logarithms arise at order higher than $\as^3$ due to the inclusion of
the bottom mass effects.  In the second approach, that we call RMB
(standing for Resummed $m_b$), we assume that all what is present at
order $\as^3$ due to bottom mass effects exponentiates. In order to do
this, for every Higgs production event, we multiply the \MiNLO{}
Sudakov form factor by
\begin{equation}\label{eq:DeltaMB}
\Delta_{m_b}(p_{t,H}, y_H) = 
\exp\left[-\int_{p'_{t,H}> p_{t,H}} \left(\frac{R_{tb}}{B_{tb}}-
\frac{R_t}{B_t}\right)\;\mathd \PhiR'\right]\,.
\end{equation}
Since
\begin{equation}\label{eq:DeltaMB}
\int_{p'_{t,H}> p_{t,H}}
\frac{R_t}{B_t}\,\mathd \PhiR' \approx
\int_{p'_{t,H}> p_{t,H}}
\frac{R_{m_t=\infty}}{B_{m_t=\infty}}\,\mathd \PhiR'\;,
\end{equation}
we notice that if we use the Sudakov form factor of
eq.~(\ref{eq:powhegsud}), evaluated in the infinite quark-mass limit,
instead of the \MiNLO{} Sudakov, the $\Delta_{m_b}$ factor turns it
into the Sudakov of eq.~(\ref{eq:powhegsud}) including bottom mass
effects. Thus, this second procedure should yield results that are
closer to those of the \gghmass{} generator.

\subsection{Quark-mass effects in \HNNLO{}}

Quark-mass effects have been implemented recently at NLO in the
version 2 of the \HNNLO{}~\cite{Grazzini:2013mca} code. For the NNLO
predictions used in this paper, we set {\tt approxim}=2 which
corresponds to having exact top and bottom mass dependence at NLO and
no approximate mass effects at NNLO.

\section{Numerical study}
We begin by considering the \HJ{}-\MiNLO{} generator, with the
quark-mass effects included at different levels.  We consider Higgs
production at the 8 TeV LHC.  We set $\mH=125.5\;$ GeV, and use in the
following MSTW2008NNLO parton distribution
functions~\cite{Martin:2009iq}. For the heavy quark-masses we use
$m_t=172.5$ GeV and $m_b=3.38$ GeV. The choice of the bottom mass
deserves some discussion.
\subsection{The choice for the bottom mass.}
In our computation, quark-mass renormalization plays no role. Thus, we
have freedom in our choice of the quark-mass scheme that we should
use, either an on-shell scheme, or the $\MSBAR{}$ or \DR{} scheme. The
one-loop relation among the masses in the different schemes is as
follows:
\begin{eqnarray}
m_{\MSBAR}(Q) &=& m_{\rm pole}\times\left[1-\frac{\as}{\pi}
\left(\log \frac{Q^2}{m^2_{\rm pole}}+4/3\right) +{\cal O}(\alpha_s^2)\right], \\
m_{\rm DR}(Q) &=& m_{\rm pole}\times\left[1-\frac{\as}{\pi}
\left(\log \frac{Q^2}{m^2_{\rm pole}}+5/3\right)+{\cal O}(\alpha_s^2)\right].
\end{eqnarray}
Choosing $\as=0.115$ and $m_{\rm pole}=4.75$ GeV, for $Q=125.5\;$GeV
we get 3.38 and 3.32 GeV for the $\MSBAR{}$ and \DR{} schemes respectively.

One could argue that, since the bottom appears in a loop involving
momenta of the order of the Higgs mass, it should be more appropriate
to use the bottom $\MSBAR{}$ mass evaluated at the Higgs mass scale.
The following observation also supports this view.  We compute the
Higgs total cross section $\sigma_{t}^{\rm LO|NLO}$ (using the
\gghmass{} generator) including only the top contribution at LO and
NLO level.  We then compute the same cross sections, $\sigma_{tb}^{\rm
  LO|NLO}$, including also the bottom loop, in the $\MSBAR{}$ and
on-shell scheme. The results for the ratios are shown in
table~\ref{tab:schemes}.
\begin{table}[htb]
\begin{center}
\begin{tabular}{|c|c|c|}
\hline
                                              & on-shell scheme & $\MSBAR{}$ scheme \\
\hline
${\sigma_{tb}^{\rm LO}}/{\sigma_{t}^{\rm LO}}$ & 0.89 & 0.93 \\ 
\hline
${\sigma_{tb}^{\rm NLO}}/{\sigma_{t}^{\rm NLO}}$ & 0.93 & 0.94 \\ 
\hline
\end{tabular}
\end{center}
\caption{Effect of the mass schemes on the inclusive Higgs cross section
at LO and NLO level.}
\label{tab:schemes}
\end{table}
We see that using the $\MSBAR{}$ value at the LO level yields a mass
effect that is closer to the one obtained at NLO.  In the following,
we will thus use the $\MSBAR{}$ value.  On the other hand, we have
verified that changing the scheme for the top mass changes the cross
section only at the per-mille level, both at LO and NLO.
 
\subsection{Mass effects in matrix elements}
For the following plots, we only consider results at the level of the
Les Houches output, i.e. no shower effects are included.

\begin{figure}[htb]
\begin{center}
\epsfig{file=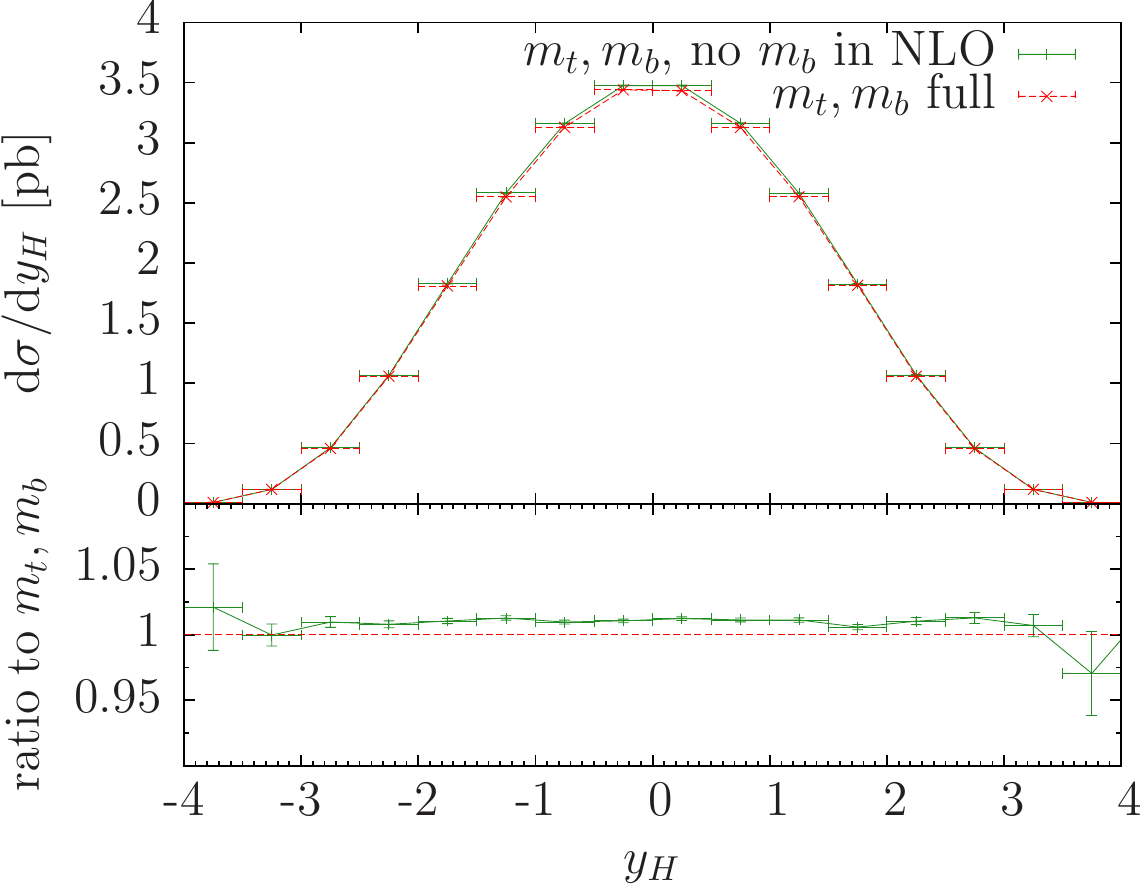,height=0.2227\textheight}
\epsfig{file=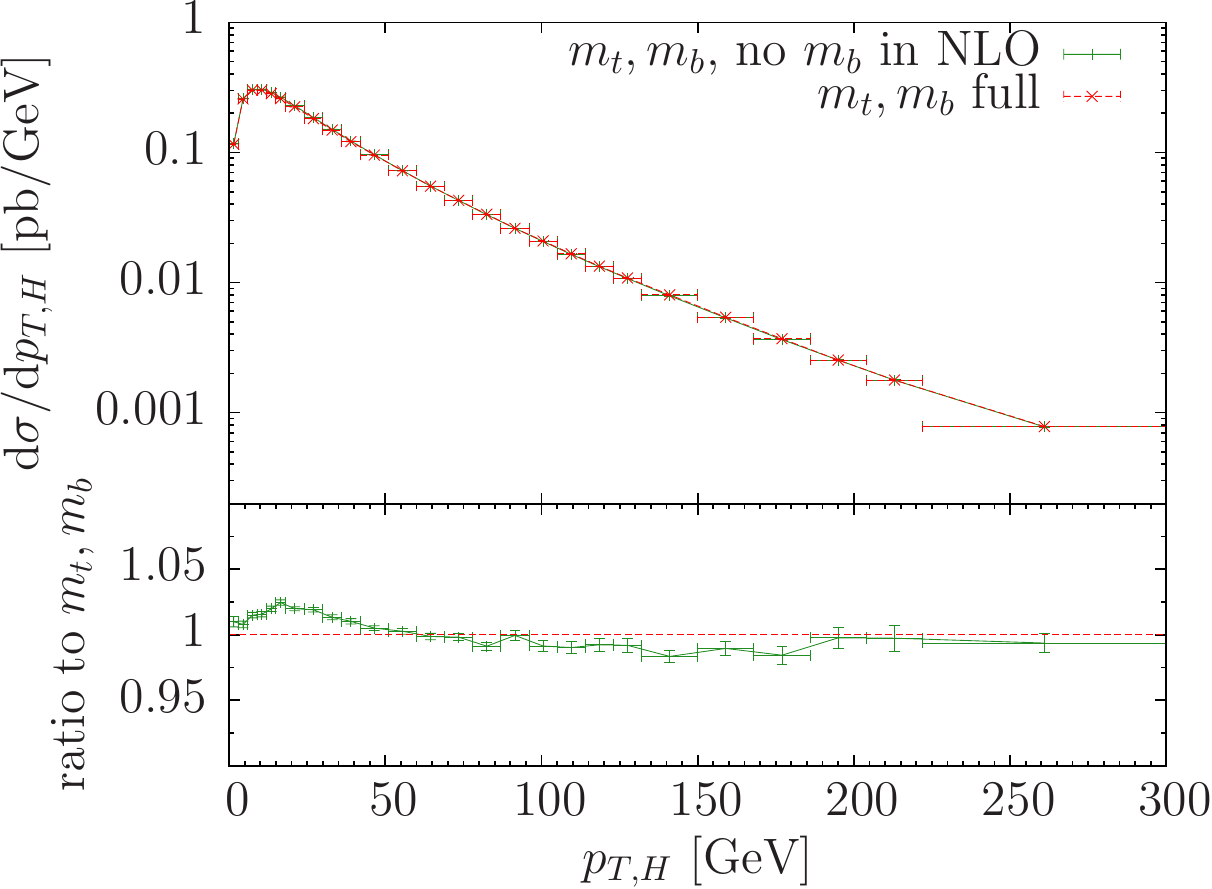,height=0.2227\textheight}
\end{center}
\caption{Rapidity and transverse momentum distributions for Higgs
  production at the 8 TeV LHC. The two lines represent the result from
  the \HJ{}-\MiNLO{} generator with bottom and top mass effects
  fully included in the matrix elements (red, dashed line),
  and the same generator without bottom mass
  effects in the matrix elements for the NLO corrections (green, solid).}
\label{fig:HJ-mtmb-no-b-in-nlo-ptH}
\end{figure}

First of all, we would like to assess the difference between the two
variants for the implementation of mass corrections in the matrix
elements, discussed in Sec.\ref{sec:meHJ}, i.e. whether the
bottom-mass correction should be only applied to the Born term, or to
the full NLO cross-section.
In figure~\ref{fig:HJ-mtmb-no-b-in-nlo-ptH}
we compare the two procedures for both the Higgs rapidity distribution
and the Higgs transverse momentum. As we can see, the differences are
at most 2\%{}. We will thus, as our default option, include also the
bottom quark in the rescaling factor for the NLO corrections.  The
other option is still provided in the public code. It can be used to
assign a systematic error associated to using an approximated
treatment of bottom-mass effects at NLO in \HJ{}.

\begin{figure}[htb]
\begin{center}
\epsfig{file=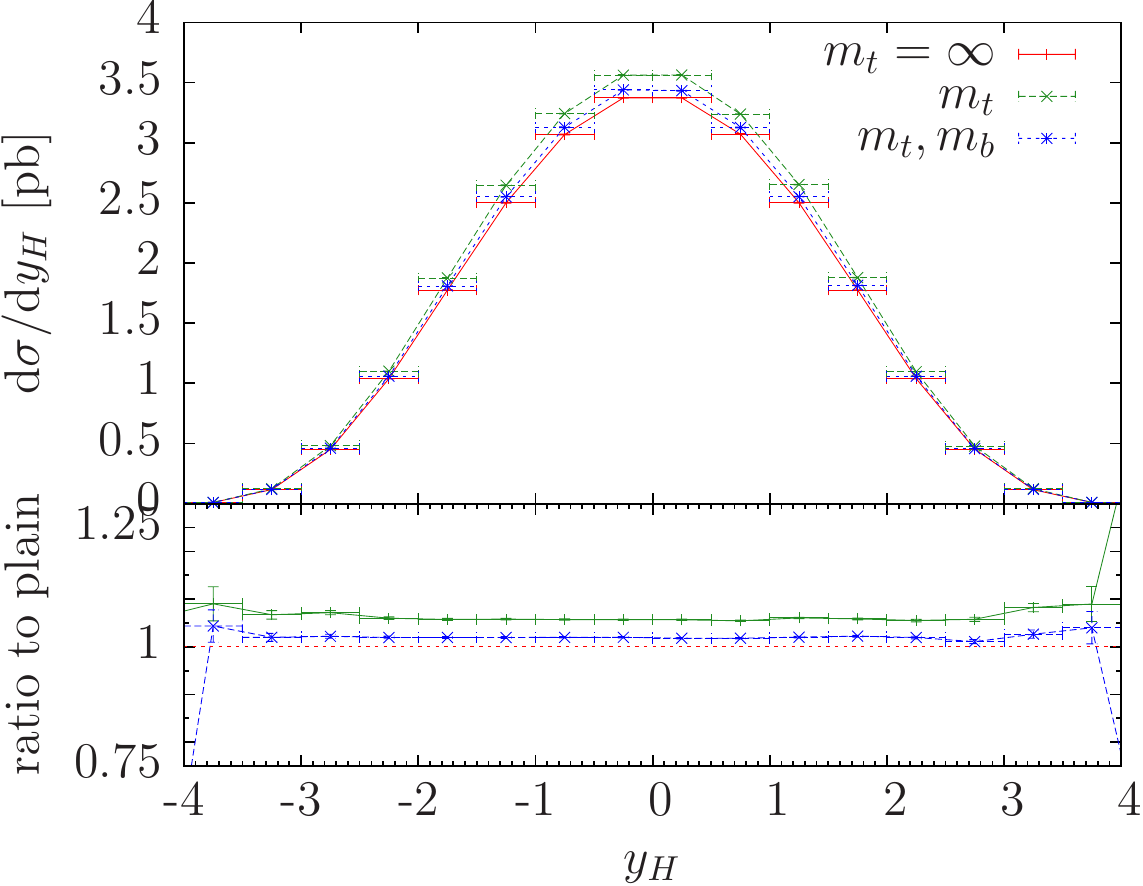,height=0.2227\textheight}
\epsfig{file=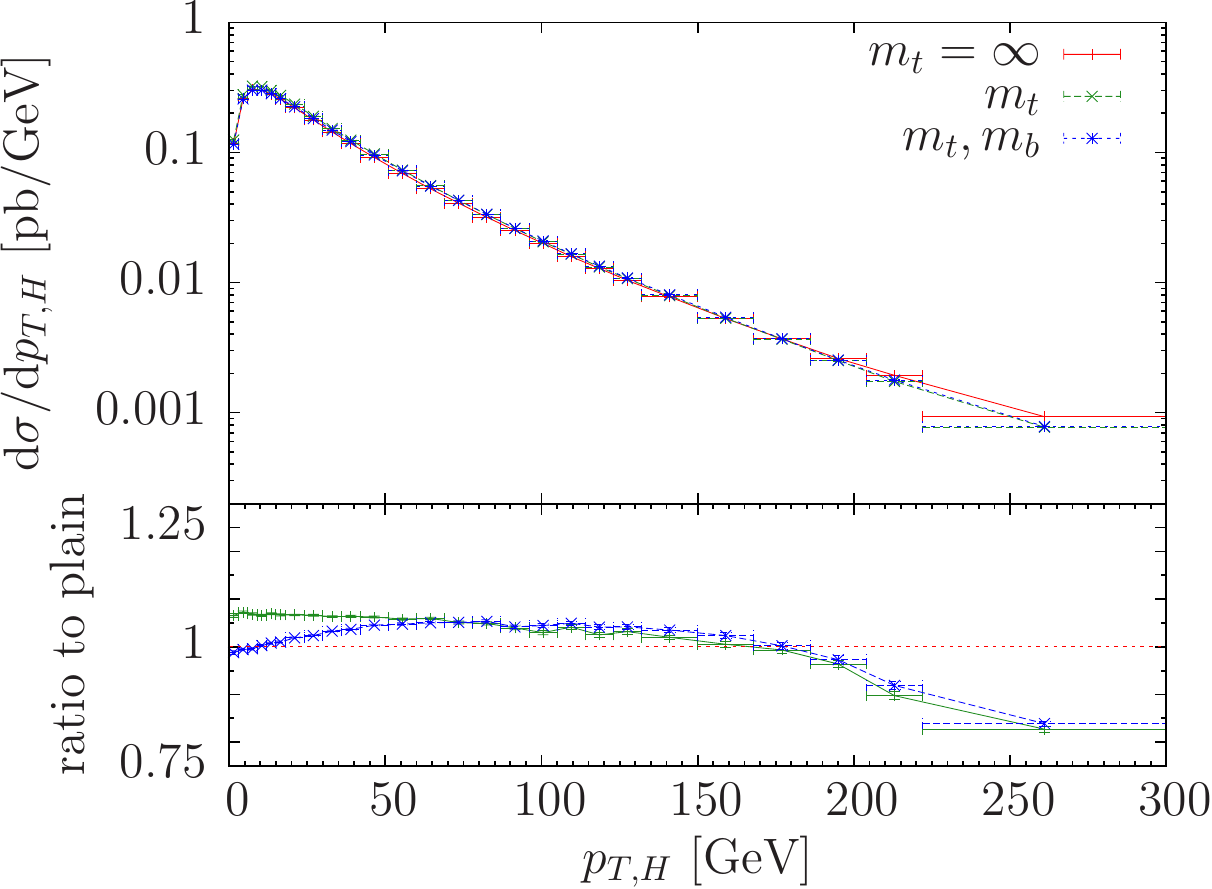,height=0.2227\textheight}
\end{center}
\caption{Rapidity and transverse momentum distributions for Higgs
production at the 8 TeV LHC. The three lines represent the
result from the plain \HJ{}-\MiNLO{} generator, and the
result from the same generator improved with either the top
mass effect alone, or with both top and bottom mass effects.}
\label{fig:HJ-0-mt-mtmb}
\end{figure}

We now compare the plain \HJ{}-\MiNLO{} generator (without any
quark-mass effects) to our new generator with mass effects included in
the matrix elements only (and no mass effects included in the \MiNLO{}
Sudakov form factor), corresponding to our MEMB option.  These are
displayed in fig.~\ref{fig:HJ-0-mt-mtmb}
for the rapidity and transverse momentum distribution of the Higgs
boson.  From the rapidity distribution, we notice that the inclusion
of the top mass alone amounts to a constant increase of about 5\%{}.
On the other hand, the inclusion of the bottom-mass effects decreases
the cross section by about the same amount, thus yielding a net
negligible effect on the total cross section.  A very similar pattern
is observed in the NLO total Higgs production inclusive cross section.
For example, ref.~\cite{Grazzini:2013mca} quotes a 6\%{} increase due
to the top mass, and a -1\%{} effect when also bottom effects are
included.
 
The similarity between the \HJ{}-\MiNLO{} and the fixed-order
inclusive NLO results for the total cross section is not trivial. In
fact, the \HJ{} matrix elements are those for Higgs production in
association with one parton at NLO, and with two partons at LO, while
the fixed-order inclusive NLO case involves the matrix elements for
the production of the Higgs without accompanying partons at NLO, plus
the production of the Higgs with one extra parton at LO. In general,
the \MiNLO{} procedure guarantees that even when the extra partons in
the \HJ{} generator are integrated out, the same accuracy of the fixed
order, inclusive NLO calculation is achieved. In the present case,
since the mass corrections in \HJ{}-\MiNLO{} are included only in an
approximate way, mass effects in the inclusive cross section do not
have full NLO accuracy. As we can see, however, the same pattern of
the NLO calculation is recovered.
From the transverse momentum spectrum we notice that, as expected, the
difference between including only the top, or both the top and the
bottom quark is concentrated at low transverse momenta, below $p_{T,H}
= 50$ GeV. At zero transverse momentum this difference is largest and
amounts to about 12\%.

\subsection{Resummed mass effects}
We still consider results at the Les Houches level, and compare the
plain \HJ{}-\MiNLO{} generator (without any quark-mass effects) to our
new generator with mass effects included in the matrix elements, and
also including the correction to the \MiNLO{} Sudakov given in
eq.~(\ref{eq:DeltaMB}), corresponding to our RMB option.

\begin{figure}[htb]
\begin{center}
\epsfig{file=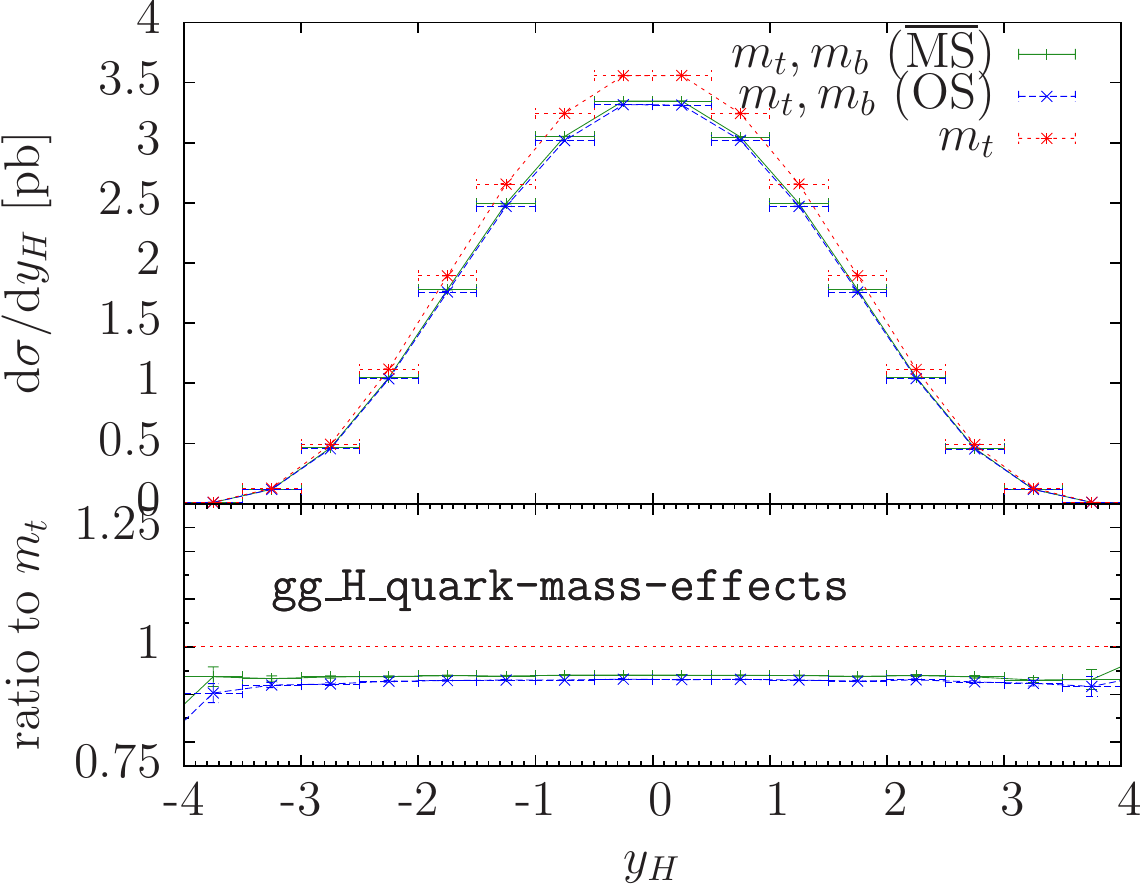,height=0.2227\textheight}
\epsfig{file=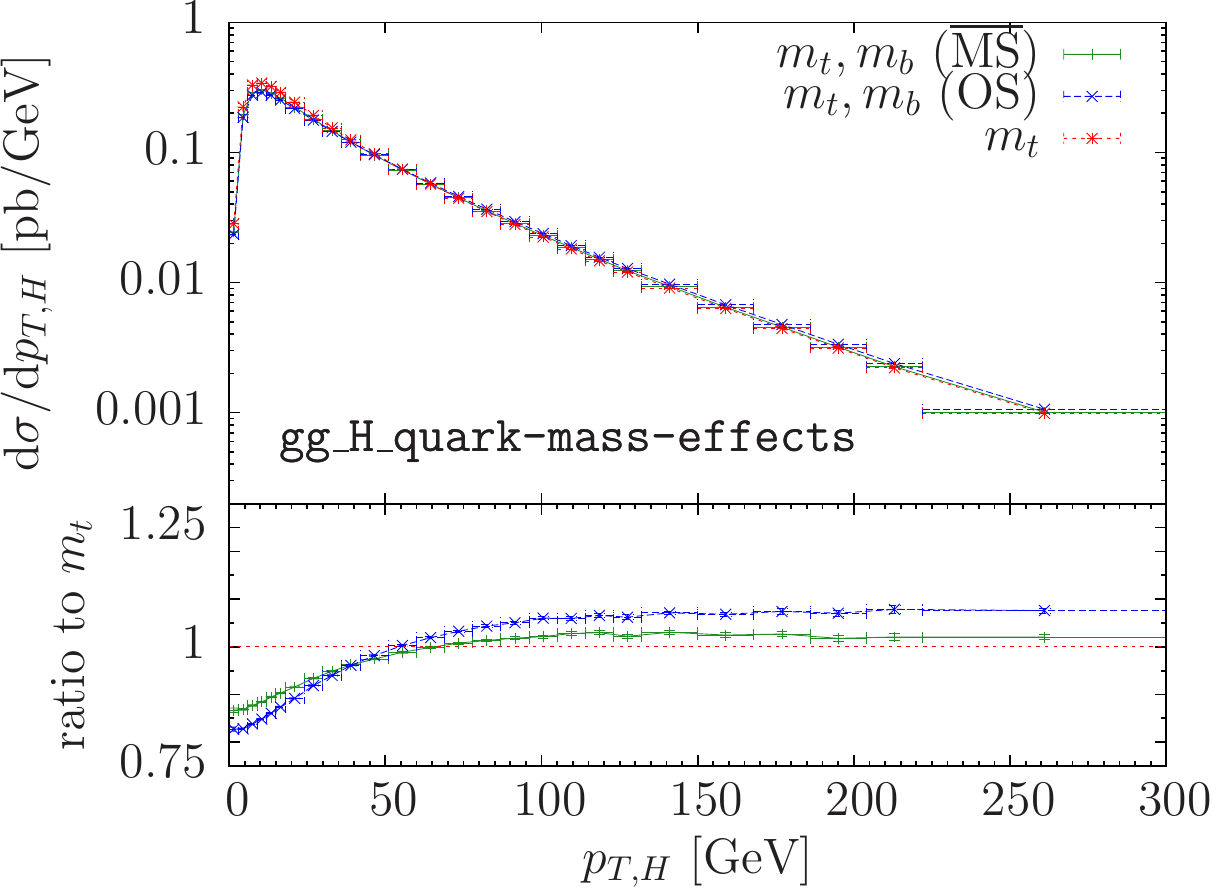,height=0.2227\textheight}
\end{center}
\caption{Rapidity and transverse momentum distributions for Higgs
  production at the 8 TeV LHC. The three lines represent the result
  from the \gghmass{} generator, when only the top loop is included,
  and when the bottom loop is added in the $\MSBAR$ and in the
  on-shell scheme.}
\label{fig:bagnaschi-os-msb}
\end{figure}

Before doing so, we remind the reader that we expect these results to
match to some extent those of the \gghmass{} generator. We thus begin
by showing in fig.~\ref{fig:bagnaschi-os-msb}
 results obtained with the \gghmass{}, when only the top loop is
 considered, in comparison to the case when also the bottom loop is
 included. In this last case, we display results in both schemes, the
 onshell and $\MSBAR$ scheme.  As explained before, in our opinion the
 latter choice is preferable in this case with respect to the on-shell
 scheme.  We see that, in the $\MSBAR$ scheme, in the rapidity
 distribution the difference is very small.  This is due to the fact
 that scheme compensation takes place in inclusive quantities,
 i.e. the explicit modification of the virtual contribution due to the
 scheme change compensates the variation induced by the change in the
 mass parameter.  This compensation does not take place in the
 transverse momentum distribution, that can be considered a leading
 order quantity, mostly affected directly by the change in the bottom
 mass.  Furthermore, there is an explicit $m_b$ dependence in the
 Sudakov exponent, where no scheme compensation occurs. In the onshell
 scheme we find more pronounced differences, both at low and high
 transverse momenta.

\begin{figure}[htb]
\begin{center}
\epsfig{file=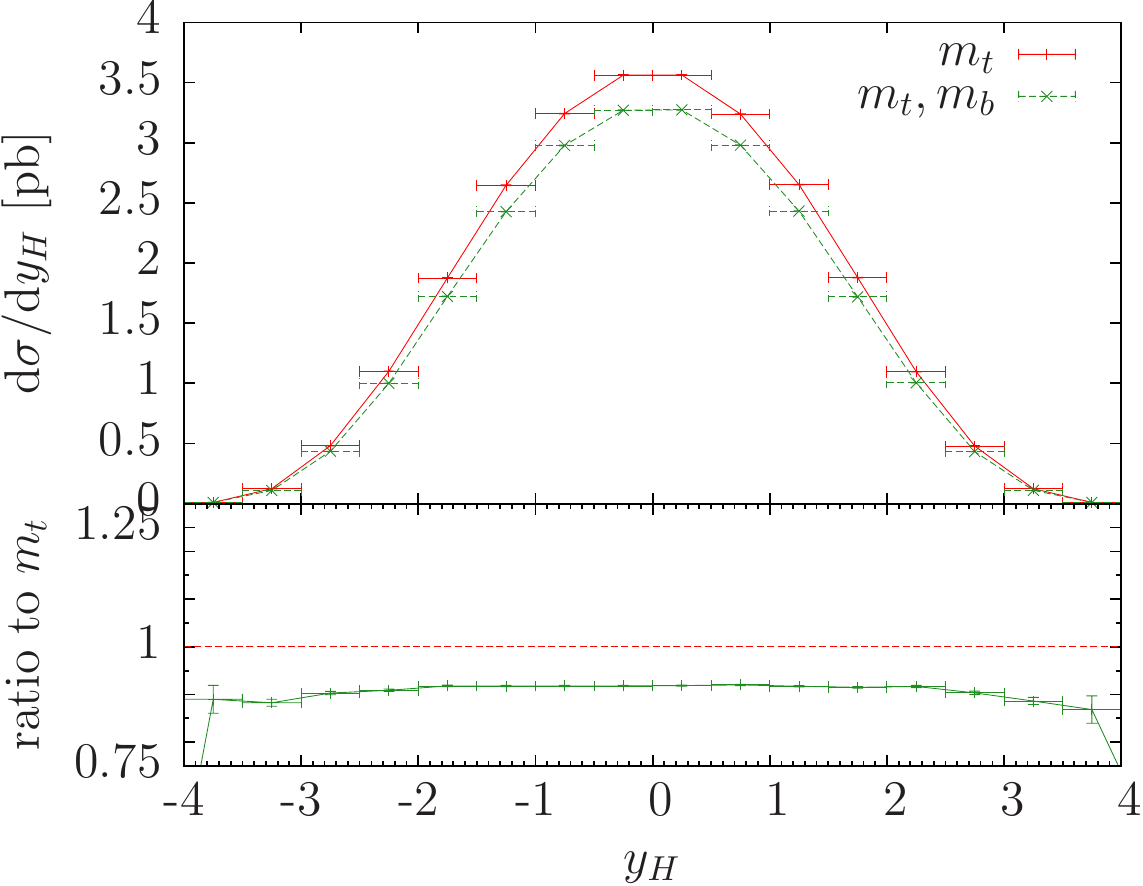,height=0.2227\textheight}
\epsfig{file=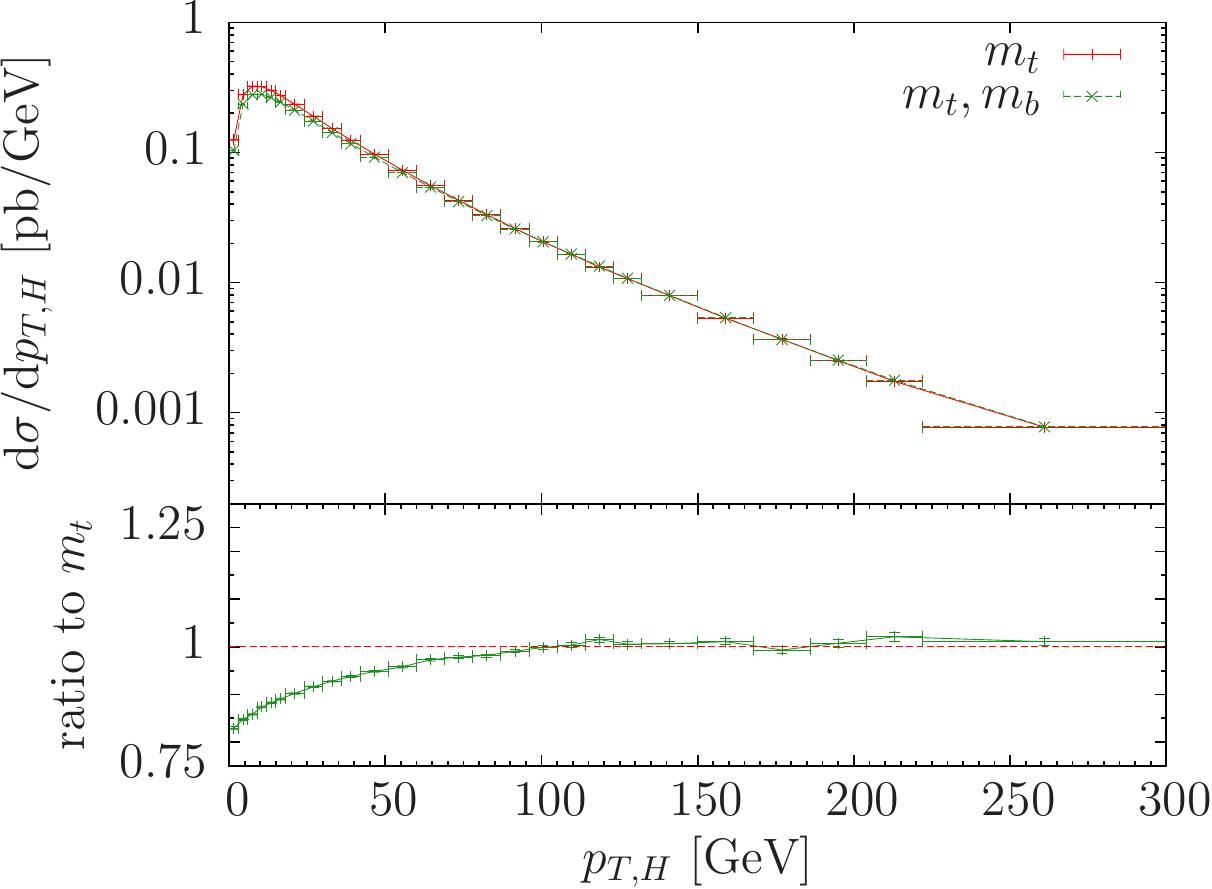,height=0.2227\textheight}
\end{center}
\caption{Rapidity and transverse momentum distributions for Higgs
  production at the 8 TeV LHC. The two lines represent the result from
  the \HJ-\MiNLO{} generator, when only the top loop is included, and
  when the bottom loop is added (in the $\MSBAR$ scheme) in our RMB
  option.}
\label{fig:HJ-RMB}
\end{figure}

We now show in fig.~\ref{fig:HJ-RMB}
the \HJ{}-\MiNLO{} results in our RMB approach (red line) compared to
including only top-mass effects (green line). As anticipated, the
bottom-mass effect is now very similar to the one displayed in
fig.~\ref{fig:bagnaschi-os-msb}.

\subsection{\NNLOPS{} results}

We now turn to our full \NNLOPS{} results. The \NNLOPS{} events are
showed with \PYTHIA{} (v.6.4.25)) with the 2011 Perugia tune ({\tt
  PYTUNE(350)}). Results include full hadronization and underlying
event effects.

\begin{figure}[htb]
\begin{center}
\epsfig{file=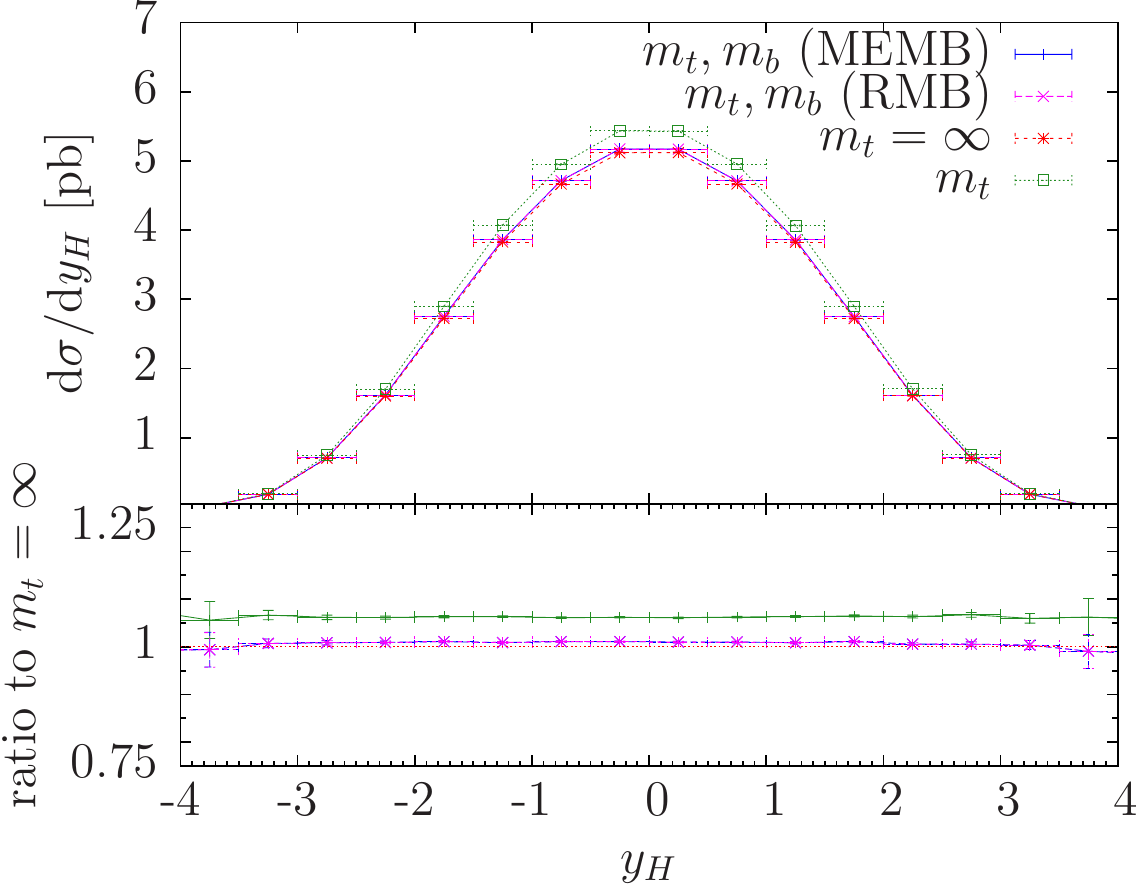,height=0.2227\textheight}
\epsfig{file=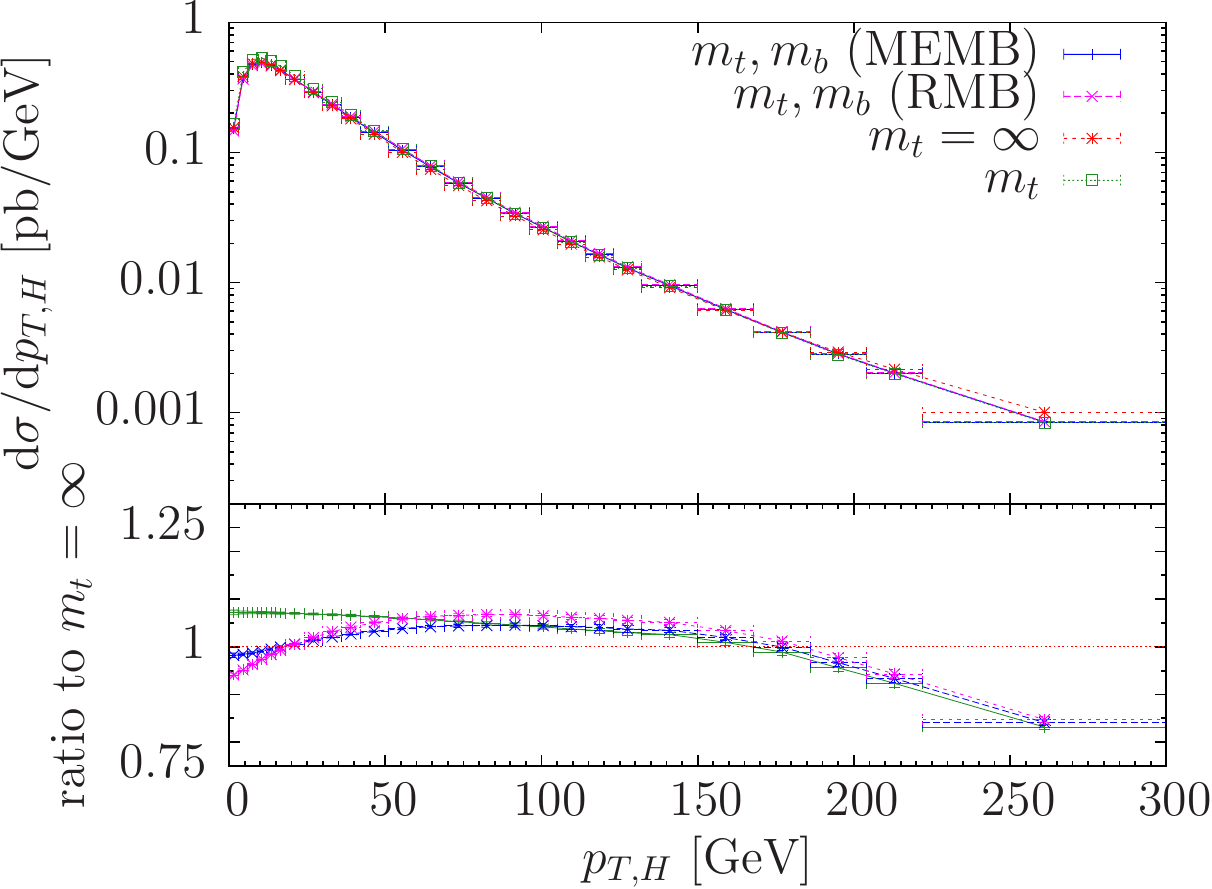,height=0.2227\textheight}
\end{center}
\caption{Rapidity and transverse momentum distributions for Higgs
  production at the 8 TeV LHC. The two lines represent the result from
  the \NNLOPS{} generator, in the large $m_t$ approximation (red),
  including only top-mass effects (green), and including top and
  bottom mass effects in the MEMB scheme (blue) or in the RMB scheme
  (magenta).}
\label{fig:NNLOPS}
\end{figure}

We start from the Higgs rapidity and transverse momentum
distributions, that are displayed in fig.~\ref{fig:NNLOPS}.

We see that, for the rapidity distribution the MEMB and RMB results
are very close to each other, and in fact very close to the large
$m_t$ result. On the contrary, results including only top loops are
about 6\% larger. This is in fact the result that one obtains at pure
NNLO level. In the transverse momentum distribution on the other hand,
we observe a difference, of up to 5\%, between the RMB and MEMB
schemes.

\begin{figure}[htb]
\begin{center}
\epsfig{file=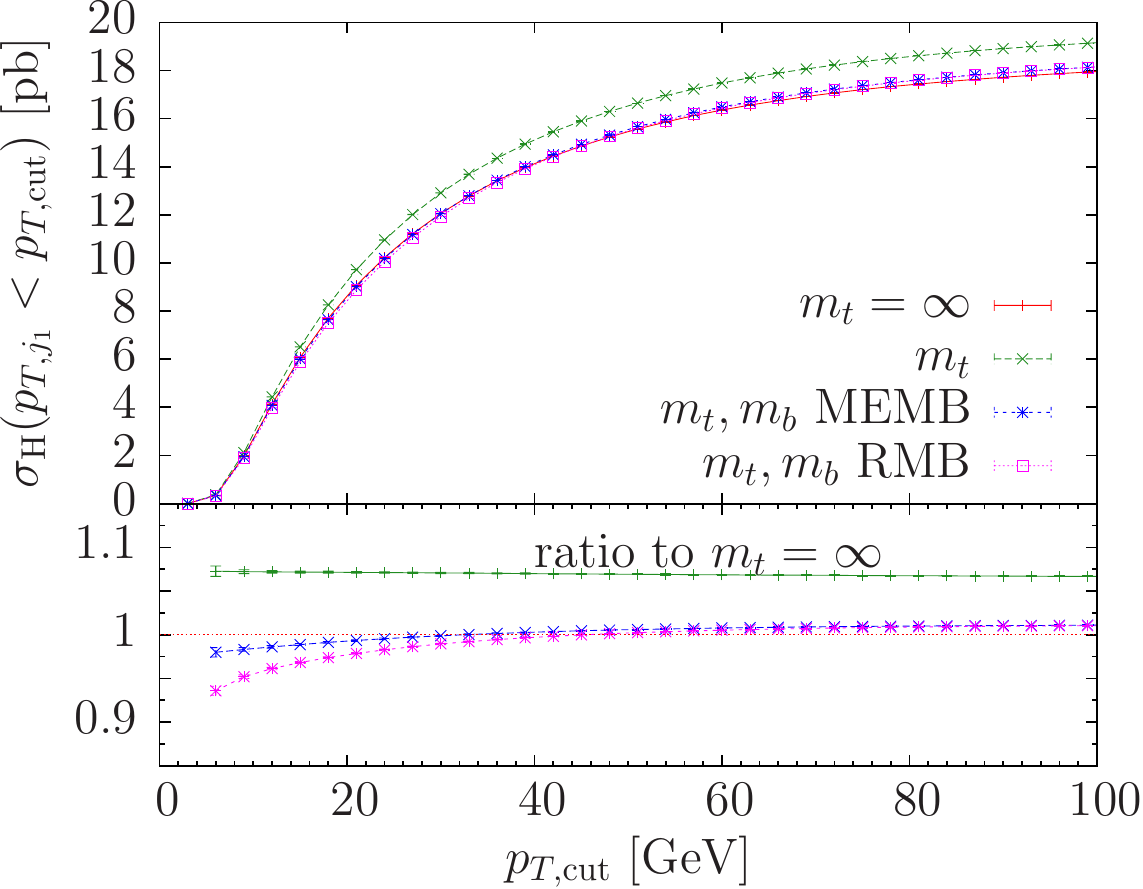,height=0.2227\textheight}
\epsfig{file=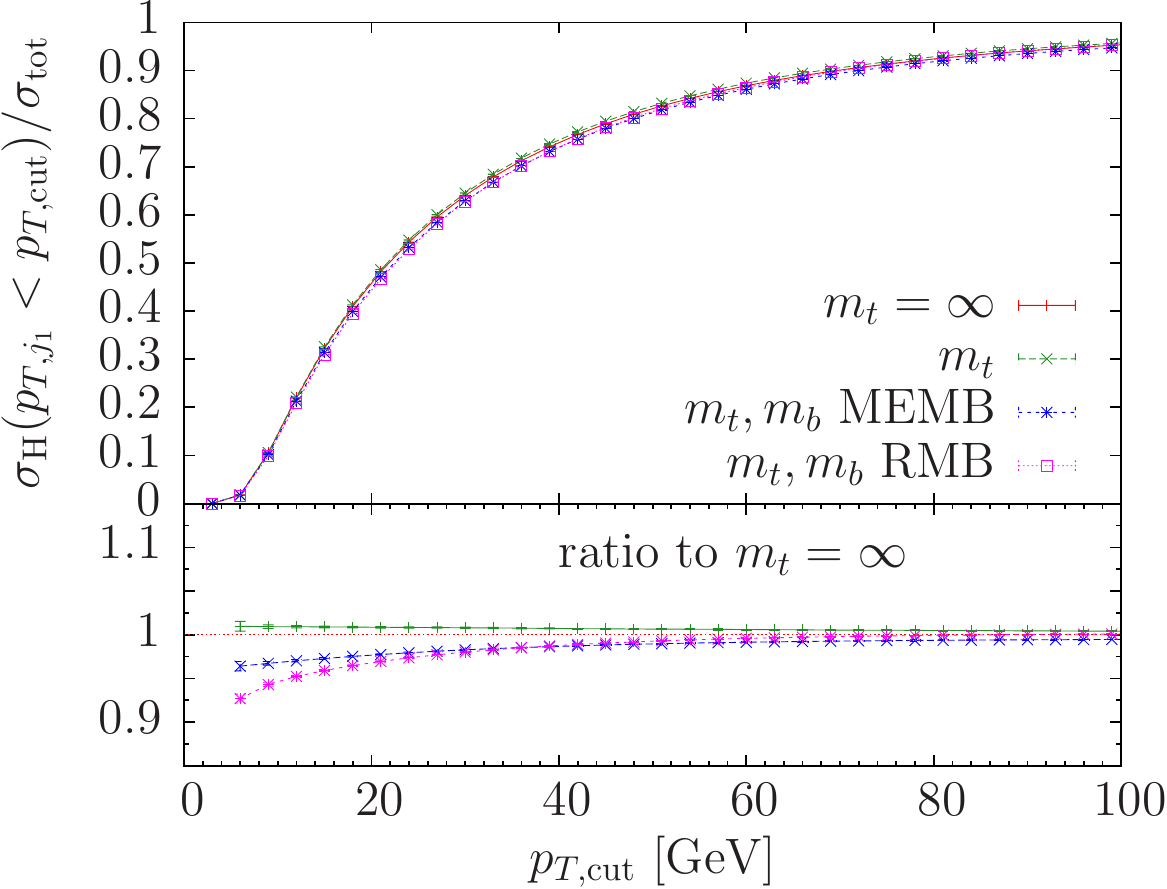,height=0.2227\textheight}
\end{center}
\caption{As in Fig.~\ref{fig:NNLOPS} but for the leading jet
  integrated cross-section (left) and jet-veto efficiency (right).}
\label{fig:NNLOPS-ptj1}
\end{figure}

In fig.~\ref{fig:NNLOPS-ptj1} we show
the four predictions for the leading jet integrated
cross-section (left) and jet-veto efficiency (right). We notice that
the difference between the MEMB and RMB schemes is very small (of the
order of 1-2\%) for values of the transverse momentum of the order of
25-30 GeV, the region of interest in Higgs studies involving a
jet-veto in ATLAS and CMS. A similar conclusion was found in
ref.~\cite{Banfi:2012jm} where a resummation for the jet-veto was
presenting including various matching procedure to estimate the
uncertainty due to the top-bottom interference contributions.
On the other hand this difference rises at lower value of the jet-veto
cut, and reaches about 5\% at $p_{\rm T,cut}= 5$ GeV.

\begin{figure}[htb]
\begin{center}
\epsfig{file=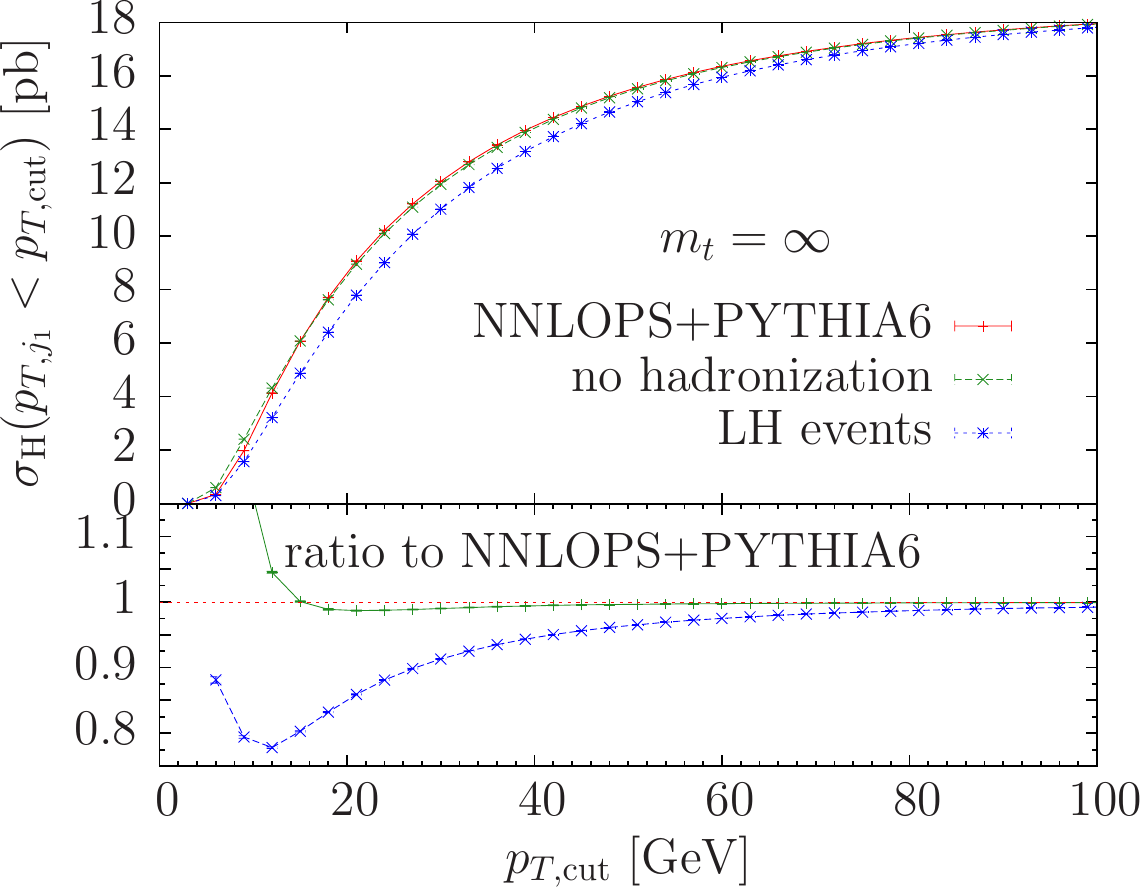,height=0.2227\textheight}
\epsfig{file=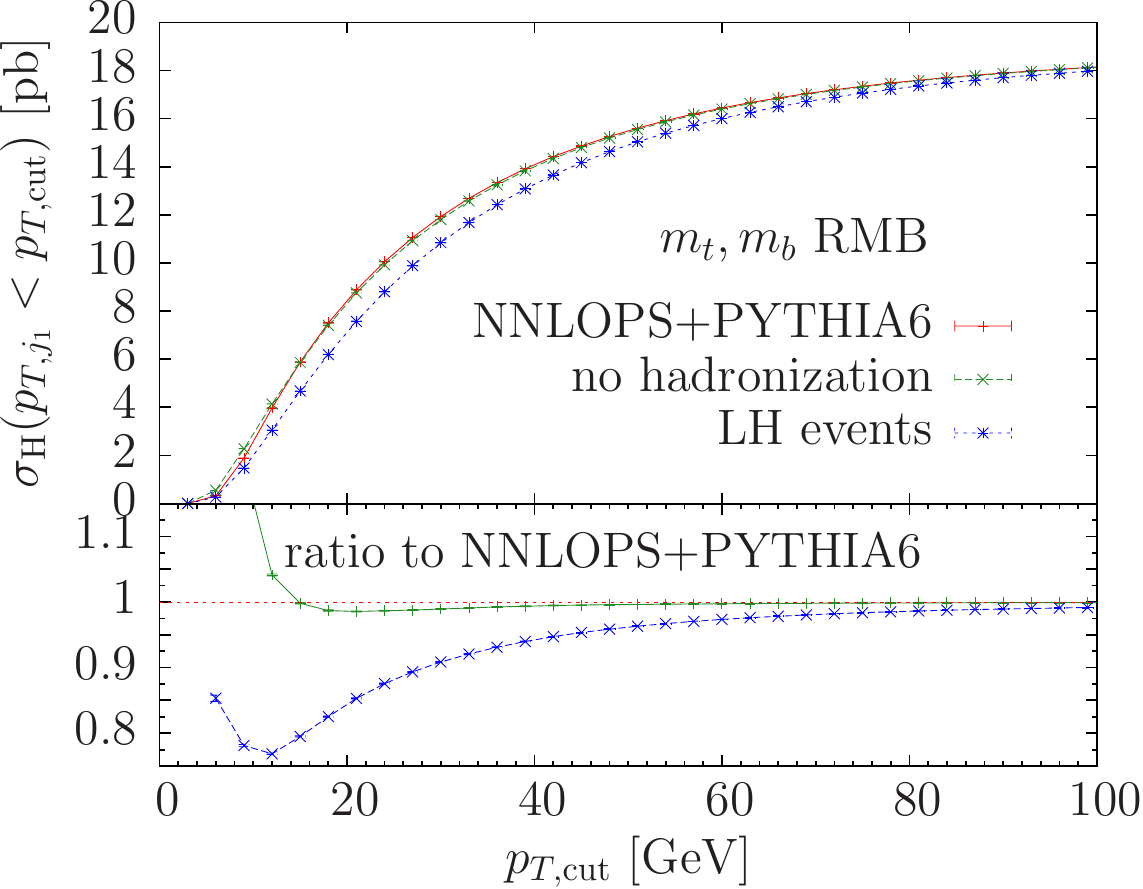,height=0.2227\textheight}
\end{center}
\caption{
Comparison of the result after shower, hadronization and inclusion
of the underlying event (red), after shower only (green),
and at the Les Houches level (blue)
for the leading jet integrated cross section. On the left, the $m_t=\infty$
result is shown, and on the right the RMB result is reported.}
\label{fig:NNLOPS-PY-LH-ptj1}
\end{figure}

Finally, in fig.~\ref{fig:NNLOPS-PY-LH-ptj1}
we show the effect of parton shower and hadronization for both the
$m_t=\infty$ case and the $m_t,m_b$, RMB result. As we can see, the
parton shower effect has a noticeable impact, while hadronization and
the underlying event do not affect this observable sensibly. In this, we
see no difference between the $m_t=\infty$ and the $m_t,m_b$, RMB case.

\section{Conclusions}

In this work, we have included finite quark-mass effects in our
\NNLOPS{} generator for Higgs production.  As far as mass-corrections
are concerned, the accuracy of our generator is that of currently
available fixed-order calculations, i.e.\ NLO accuracy.
In our procedure we have considered two possible sources of
uncertainty. The first one, has to do with the fact that we rescale
the full matrix elements with the mass effect that are computed at the
level of Higgs production in association with one jet. This procedure
may be considered safe as far as top mass-effects are concerned, where
one may argue that dominant NLO corrections, involving momenta softer
than the Higgs mass, will not affect the top-quark loop. This
assumption is unjustified when the bottom loop is involved. We thus
consider two alternatives: we either apply the mass-correction only to
the Born contribution, or to the full NLO matrix element. We find that
numerically, the two procedures lead to differences of the order of
2\%.
A second source of uncertainty has to do with the all-order inclusion
of contributions enhanced by powers of $\log(m_b/m_H)$. Little is
known about the all-order structure of these logarithms. We thus
consider two options: either we do not include them at all, or we
fully exponentiate them in the \MiNLO{} Sudakov form factor. The first
approach is similar in spirit to applying a resummation procedure by
assuming that soft gluon momenta do not affect the quark loops. The
second approach is analogous to what is implemented in the \gghmass{}
generator, where the \POWHEG{} procedure leads to full exponentiation
of the real emission cross-section including mass effects. In this
case we find differences in the low transverse momentum region of the
order of 5\% starting from the low momentum region and changing
direction at high transverse momenta. On the other hand, when
considering jet-veto distributions for typical values of the
transverse momentum cuts used by ATLAS and CMS ($p_{\rm T,cut} =
25$-$30$ GeV), we find only 1-2\% differences.

Unlike in ref.~\cite{Hamilton:2013fea} we have not reported here a
study of scale variation, since in view of the smallness of the
quark-mass effects, we do not expect to get results sensibly different
from those of reference~\cite{Hamilton:2013fea}.

The code for this generator has been made available in an update of
the \HJ{} package in the {\tt POWHEG-BOX-V2}.

\section*{Acknowledgments}
The research of GZ is supported by the Consolidator ERC Grant 614577.
We would like to thank the Galileo Galilei Institute (PN and GZ) and
CERN (PN) for hospitality while part of this work was carried out.
KH is grateful to the Institute for Particle Physics Phenomenology 
in Durham, for support in the form of an IPPP Associateship.

\appendix

\newcommand{\nocomma}{}
\newcommand{\tmop}[1]{\ensuremath{\operatorname{#1}}}
\newenvironment{enumeratenumeric}{\begin{enumerate}[1.] }{\end{enumerate}}

\section{Technical details about $b$-mass effects in the \MiNLO{} Sudakov}
\label{sec:app}

The inclusion of bottom-mass effects in the \MiNLO{} Sudakov,
following eq.~\eqref{eq:DeltaMB}, requires for each point an
additional integration over a ratio a matrix elements, for fixed
$y_H$, over the phase space region where the Higgs transverse momentum
is larger than the Higgs transverse momentum of the event,
$p_{T}$. Since such integration is performed on the fly for each
point, it is crucial that the phase space is sampled
efficiently. Furthermore, we will use an approximate expression of the
matrix elements appearing in eq.\eqref{eq:DeltaMB}. In
Appendix~\ref{app:phasespace} we give all details about the phase
space evaluation, while in Appendix~\ref{app:approxME} we illustrate
the approximations done on the matrix elements in
eq.~\eqref{eq:DeltaMB}.

\subsection{Phase space}\label{app:phasespace}

Here we study the phase space integration of the matrix element for
the production of a Higgs boson plus one light parton, under the
constraint that the Higgs boson has fixed rapidity $y_H$, and that its
transverse momentum is larger than a given lower limit $p_T$.

We denote by $m_H$ the Higgs mass, $k_H = (E,\vec{k})$ its four
momentum, by $l_g = (l,\vec{l})$ the light-parton momentum and define
$q = k_H+l_g$. The three-vectors $\vec{l}$ and $\vec{k}$ are defined
in the partonic center of mass (CM) frame.  We have also defined
$k=\vec{k}$ and $l=\vec{l}$.  The two body phase space in the rest
frame of the Higgs-light-parton system is then given by:
\begin{eqnarray}
d \Phi_2 & = &   \frac{d^3 \vec{l}}{2 l ( 2 \pi)^3} 2 \pi \delta ( ( q - l_g)^2 - m_H^2)\nonumber\\
 &=& 
  \frac{d^3 \vec{l}}{2 l ( 2 \pi)^2} \delta ( q^2 - m_H^2 - 2 q^0 l) = \frac{d^2
  l_T d l_l}{2 l ( 2 \pi)^2} \delta \left( q^2 - m_H^2 - 2 q^0 \sqrt{l_T^2 +
  l_l^2 } \right)\nonumber \\
& = & 
  \sum_{\pm} \frac{1}{16 \pi}  \frac{\theta ( l^2 - l_T^2)}{\sqrt{s} \sqrt{l^2
  - l_T^2}} d l_T^2 = \sum_{\pm} \frac{1}{16 \pi}  \frac{\theta ( k^2 -
  k_T^2)}{\sqrt{s} \sqrt{k^2 - k_T^2}} d k_T^2\,,
\end{eqnarray}
where $s = q^2$ and where we have used the property $\vec{k} =
-\vec{l}$. The $\pm$ sum refers to the two possible signs of the $l_l$
integral (the suffix $l$ stands here for longitudinal).

The full phase space integral, to be multiplied by parton distribution
functions and partonic cross section, including the constraint that
the Higgs transverse momentum is larger than $p_T$, is then
\begin{equation}
  \int \mathd k_T^2 \sum_{\pm} \frac{1}{16 \pi}  \frac{1}{\sqrt{s} \sqrt{k^2 -
  k_T^2}} \int \mathd y_{\tmop{cm}} \mathd \tau \; \delta ( \pm
  y_{\tmop{cm}}^H - ( y^H - y_{\tmop{cm}})) \theta ( k^2 - k_T^2) \theta (
  k_T^2 - p_T^2)\,,
\end{equation}
where $y^H$ is the Higgs rapidity in the laboratory frame and
$y_{\tmop{cm}}^H$ the Higgs rapidity in the CM frame.  The above
expression is equivalent to
\begin{equation}
  \int \mathd k_T^2 \frac{1}{16 \pi}  \frac{1}{\sqrt{s} \sqrt{k^2 - k_T^2}}
  \int \mathd y_{\tmop{cm}} \mathd \tau \; \delta (y_{\tmop{cm}}^H - |y^H -
  y_{\tmop{cm}} |) \theta ( k^2 - k_T^2) \theta ( k_T^2 - p_T^2)\,.
\end{equation}
The Higgs rapidity in the CM frame can be written as 
\begin{equation}
  y_{\tmop{cm}}^H = \frac{1}{2} \log \frac{E + \sqrt{k^2 - k_T^2}}{E -
  \sqrt{k^2 - k_T^2}}\,,
\end{equation}
where $E, k$ are the energy and
momentum of the Higgs in the partonic CM frame: 
\begin{equation}
  E = \frac{s + m_H^2}{2 \sqrt{s}} = m_H \frac{1 + z}{2 \sqrt{z}},
  \hspace{1em} k = \frac{s - m_H^2}{2 \sqrt{s}} \; = m_H \frac{1 - z}{2
  \sqrt{z}},
\end{equation}
and $s = m_H^2 / z$. Furthermore we define 
\begin{equation}
  \tau = \frac{s}{S} = \frac{m_H^2}{S z}\,.
\end{equation}
In order to solve for $k_T^2 $ in the delta function we get
\begin{equation}
  \sqrt{k^2 - k_T^2} = E \tanh \left | y_H - y_{\tmop{cm}} \right|\,,
\end{equation}
and
\begin{equation}
  k_T^2 = k^2 - E_{}^2 \tanh^2 ( y_H - y_{\tmop{cm}}) \; .
\end{equation}
Since we must have $p_T < k_T < k$, from the theta functions we get the
constraints
\begin{equation}
  p_T^2 < k^2 - E_{}^2 \tanh^2 ( y_H - y_{\tmop{cm}}) < k^2\,,
\end{equation}
that is to say
\begin{equation}
  k^2 - p_T^2 > E_{}^2 \tanh^2 ( y_H - y_{\tmop{cm}})\,,
\end{equation}
or
\begin{equation}
  \Delta y \equiv \tmop{atanh} \sqrt{\frac{k^2 - p_T}{E_{}^2}^2 } =
  \frac{1}{2} \log \left( \frac{E + \sqrt{k^2 - p_T^2}}{E - \sqrt{k^2 -
  p_T^2}} \right) > | y_H - y_{\tmop{cm}} |\,.
\end{equation}
The jacobian for the delta function integration is
\begin{equation}
  \left| \frac{\partial y_{\tmop{cm}}^H}{\partial k_T^2} \right|^{- 1} =
  \frac{2}{E} ( E^2 - k^2 + k_T^2) \sqrt{k^2 - k_T^2} = \frac{2 ( m_H^2 + k_T^2
  ) \sqrt{k^2 - k_T^2}}{E}\,.
\end{equation}
Our phase space integral becomes
\begin{equation}
  \int \mathd \tau \int_{\frac{1}{2} \log \tau}^{- \frac{1}{2} \log \tau}
  \mathd y_{\tmop{cm}}  \frac{1}{8 \pi}  \frac{E^2 - k^2 + k_T^2}{E \sqrt{s}} 
  \; \theta ( \Delta y - | y_H - y_{\tmop{cm}} |) \theta ( k^2 - p_T^2)\,,
\end{equation}
which is equivalent to 
\begin{equation}
  \int_{\tau_0}^1 \mathd \tau \int_{\frac{1}{2} \log \tau}^{\frac{1}{2} \log 1
  / \tau} \mathd y_{\tmop{cm}}  \frac{z}{4 \pi}  \frac{m_H^2 + k_T^2}{m_H^2 (
  1 + z)}  \; \theta ( k^2 - E_{}^2 \tanh^2 ( y_H - y_{\tmop{cm}}) - p_T^2)\,,
\end{equation}
with $\tau_0 = ( p_T + M_T)^2 / S$, with $M_T = \sqrt{p_T^2 + m_H^2}$. This
limit arises from the requirement:
\begin{equation}
  k = \frac{s - m_H^2}{2 \sqrt{s}} > p_T,
\end{equation}
that implies
\begin{equation}
  s - 2 \sqrt{s} p_T - m_H^2 > 0 \Rightarrow \left( \sqrt{s} - ( p_{_T} + M_T)
  \right) \left( \sqrt{s} - ( p_T - M_T) \right) > 0,
\end{equation}
which in turn implies
\begin{equation}
  s > ( p_T + M_T)^2 .
\end{equation}
The theta function
\begin{equation}
  \theta ( k^2 - E_{}^2 \tanh^2 ( y_H - y_{\tmop{cm}}) - p_T^2)
\end{equation}
also implies some restrictions on $y_{\tmop{cm}}$, that we now work out. It
can be written as
\begin{equation}
  \tmop{atanh} \sqrt{\frac{k^2 - p_T^2}{E^2} } > |y_H - y_{\tmop{cm}} |,
\end{equation}
or
\begin{equation}
  y_H + \Delta > y_{\tmop{cm}} > y_H - \Delta,
\end{equation}
with
\begin{equation}
  \Delta = \tmop{atanh} \sqrt{\frac{k^2 - p_T^2}{E^2} } .
\end{equation}
Combined with the integration limits for $y_{\tmop{cm}}$, this yields
\begin{equation}
  \max \left( - \frac{1}{2} \log \frac{1}{\tau} \nocomma \,, \nocomma y_H -
  \Delta \right) < y_{\tmop{cm}} < \min \left( \frac{1}{2} \log \frac{1}{\tau}
  \nocomma \,, \nocomma y_H + \Delta \right) .
\end{equation}
The range is not empty if
\begin{eqnarray*}
  - \frac{1}{2} \log \frac{1}{\tau} & < & y_H + \Delta,\\
  y_H - \Delta & < & \frac{1}{2} \log \frac{1}{\tau},
\end{eqnarray*}
that can be combined into
\begin{equation}
  \Delta > - \frac{1}{2} \log \frac{1}{\tau} + |y_H | .
\end{equation}
Taking the hyperbolic tangent of both sides we get
\begin{equation}
  \sqrt{\frac{k^2 - p_T^2}{E^2}} > \frac{\tau e^{2 |y_H |} - 1}{\tau e^{2 |y_H
  |} + 1} = \frac{\tau - e^{- 2 |y_H |}}{\tau + e^{- 2 |y_H |}} .
\end{equation}
Let us define $\eta = e^{- 2 |y_H |}$; we have the inequality
\begin{equation}
  \sqrt{\frac{k^2 - p_T^2}{E^2}} > \frac{\tau - \eta}{\tau + \eta},
\end{equation}
or
\begin{equation}
  \sqrt{\frac{( \tau S - m_H^2)^2 - 4 S p^2_T \tau}{( \tau S + m_H^2)^2}} >
  \frac{\tau - \eta}{\tau + \eta},
\end{equation}
that is easily solved to yield either $\tau < \eta$, or
\begin{equation}
  \tau > \tau_1 = \frac{( \eta S - m_H^2) \sqrt{\eta S M_T^2} + \eta S
  p_T^2}{S ( \eta S - M_T^2)},
\end{equation}
or
\begin{equation}
  \tau < \tau_2 = \frac{- ( \eta S - m_H^2) \sqrt{\eta S M_T^2} + \eta S
  p_T^2}{S ( \eta S - M_T^2)} . \label{eq:tauless}
\end{equation}
Setting $\eta = x^2$, and using $S = E^2$ and $m_H^2 = M_T^2 - p_T^2$, we get (using an
algebraic manipulation code)
\begin{equation}
  \tau_1 = \frac{x ( E M_T x + p_T^2 + M_T^2)}{E ( E x + M_T)},
\end{equation}
\begin{equation}
  \tau_2 = - \frac{x ( E M_T x - p_T^2 + M_T^2)}{E ( E x + M_T)} < 0,
\end{equation}
so that only $\tau_1$ is retained, and we need to satisfy the conditions:
either $\tau > \tau_0$ and $\tau < \eta$, or $\tau > \tau_0$ and $\tau >
\tau_1$. We also find:
\begin{equation}
  \tau_1 - \eta = - x \frac{( E x - p_T - M_T) ( E x + p_T - M_T)}{E ( E x -
  M_T)} = - x \left( x - \sqrt{\tau_0} \right) \frac{E x + p_T - M_T}{E x -
  M_T}, \label{eq:tau1meta}
\end{equation}
and
\begin{equation}
  \tau_1 - \tau_0 = \frac{M_T \left( x - \sqrt{\tau_0} \right)^2}{E x - M_T}
  \label{eq:tau1mtau0}
\end{equation}
We notice that we must have $E x - M_T > 0$ in all cases, otherwise
the Higgs with transverse mass $M_T$ and rapidity $y_H$ would be
inconsistent with the total incoming energy. Hence $\tau_1 > \tau_0$,
and we get only two cases:
\begin{enumeratenumeric}
  \item $\tau_0 < \eta$: in this case the right hand side of eq.
    (\ref{eq:tau1meta}) is negative. In fact, $E \sqrt{\tau_0} - M_T$
    is obviously positive, and it remain positive if we replace
    $\sqrt{\tau_0}$ with $x$, since $x > \sqrt{\tau_0}$. So the last
    factor on the right hand side of eq. (\ref{eq:tau1meta}) is
    positive, and the right hand side is negative. Therefore $\tau_1 <
    \eta$, and the range of integration in $\tau$ is $\tau_0 < \tau <
    1$, because $\tau > \eta$ also implies $\tau > \tau_1$.
  
  \item $\tau_0 > \eta$: in this case eq. (\ref{eq:tau1mtau0}) guarantees that
  $\tau_1 > \tau_0$, and we have $\tau_1 > \tau_0 > \eta$. The range of
  integration is $\tau_1 < \tau < 1$. The condition $\tau_1 < 1$ is the
  remaining requirement of consistency for the assigned $y_H$ and $p_T$
  values.
\end{enumeratenumeric}

\subsection{Approximate matrix elements in the \MiNLO{} Sudakov}\label{app:approxME}

Here we record the approximate matrix elements used for computing our
finite quark-mass corrections to the \MiNLO{} Sudakov form factor.
For what concerns quark-mass effects in the partonic Higgs
boson-plus-one parton sub-processes, all other components entering the
construction of our \NNLOPS{} event generator utilize the exact,
leading order, $2\rightarrow2$ matrix elements. Since the
$qq\rightarrow Hg$ matrix element contains no enhanced terms
proportional to $1/p_{T}^{2}$, it cannot give rise to corrections to
the Sudakov form factor, hence it does not feature here.

The pure $b$-quark loop mediated contributions to the $p_{T}$ spectrum
are greatly suppressed by a factor $m_{b}^{2}/m_{H}^{2}$ relative to
those owing to the interference of the $t$- and $b$-loop mediated
amplitudes, hence, our first approximation has been to neglect the
former. Our second simplification is to use the relatively simpler
small quark-mass limit of the scalar loop integrals given in
ref.~\cite{Baur:1989cm} for the $b$-loop amplitudes, while using the
large-$m_{t}$ limit for the top-loop amplitudes. The small quark-mass
limit of ref.~\cite{Baur:1989cm} is defined by taking $m_{b}$ as being
small compared to all other kinematic invariants in the process.

We have validated the following approximate matrix elements
numerically in various ways. To begin with we have compared them
point-wise in phase space to the exact matrix elements. In the latter
case the evaluation of the \emph{integrand} in the correction to the
Sudakov exponent, modulo convolution with the PDFs, with the
approximate matrix elements agrees with its exact analogue to well
within $20$\% for $20\lesssim p_{T}\lesssim150\,\mathrm{GeV}$.  For
$qg\rightarrow Hg$ and $gq\rightarrow Hq$ channels the approximate
matrix elements are particularly effective, and agreement in that case
is better than 10\% for $10\ll p_{T}\ll300\,\mathrm{GeV}$. Since all
of our matrix element approximations involve the small-$m_{b}$ limit,
they can all be expected to breakdown for $p_{T}\rightarrow m_{b}$,
however, to perform the proposed correction to the Sudakov form
factor, which involves the cumulant in $p_{T}$ starting at its maximum
value, this breakdown is unimportant. We have also tested the
approximate matrix elements against their exact counterparts in the
evaluation of the full correction to the Sudakov form factor itself,
$\Delta_{m_{b}}\left(p_{T},\, y_{H}\right)$.  Due to the fact this
quantity involves an integral over $p_{T}$ up to its maximum
attainable value, and the convolution of the matrix elements with the
PDFs --- marginalising the contribution of the high $p_{T}\gtrsim
m_{t}$ region, in which the $gg\rightarrow Hg$ approximation loses
quality --- we find that results for $\Delta_{m_{b}}\left(p_{T},\,
y_{H}\right)$ obtained using the approximate matrix elements are
almost indistinguishable from those found using the exact matrix
elements.

\subsubsection{Preliminaries}

The exact spin-colour averaged\emph{ }Born amplitude squared,
including finite top- and bottom-quark-mass effects,
$\mathcal{M}_{gg\rightarrow H}^{\left[m_{t},m_{b}\right]}$, and its
large-$m_{t}$ limit, $\mathcal{M}_{gg\rightarrow
  H}^{\left[\infty,0\right]}$, are given by,
\begin{equation}
\mathcal{M}_{gg\rightarrow H}^{\left[\infty,0\right]}=\frac{G_{F}m_{H}^{4}}{288\sqrt{2}}\frac{\alpha_{{\scriptscriptstyle \mathrm{S}}}^{2}}{\pi^{2}}\,\qquad\mathrm{and}\qquad\mathcal{M}_{gg\rightarrow H}^{\left[m_{t},m_{b}\right]}=\mathcal{F}\,\mathcal{M}_{gg\rightarrow H}^{\left[\infty,0\right]}\,,\label{eq:DR-M_B-and-M_B-infty-1}
\end{equation}
where, 
\begin{equation}
\mathcal{F}=\left|\sum_{q}\frac{3}{2\tau_{q}^{2}}\left(\tau_{q}+\left(\tau_{q}-1\right)f\left(\tau_{q}\right)\right)\right|^{2}\,,\qquad f\left(\tau_{q}\right)=\left\{ \begin{array}{ll}
\arcsin^{2}\sqrt{\tau_{q}} & \quad\tau_{q}\le1\,,\\
-\frac{1}{4}\left[\log\frac{1+\sqrt{1-1/\tau_{q}}}{1-\sqrt{1-1/\tau_{q}}}-i\pi\right]^{2} & \quad\tau_{q}>1\,,
\end{array}\right.\label{eq:F_0-formula-and-f_tau-formula-1}
\end{equation}
with $\tau_{q}=\frac{1}{4}m_{H}^{2}/m_{q}^{2}$ and $m_{q}$ being
the mass of the quark ($q=t,\, b$). In the large-$m_{t}$-small-$m_{b}$
limit, 
\begin{eqnarray}
\underset{{\scriptstyle m_{b}\rightarrow0}}{\lim_{m_{t}\rightarrow\infty}}\mathcal{F} & \rightarrow & 1+\frac{3}{4}\,\tau_{b}^{-1}\,\mathcal{F}_{tb}+\mathcal{O}\left(\tau_{b}^{-2}\right)\,,\qquad\mathcal{F}_{tb}=4+\pi^{2}-\log^{2}\left(4\tau_{b}\right)\,.\label{eq:F_0-formula-squared}
\end{eqnarray}
In eq.~\eqref{eq:F_0-formula-squared} the leading term
$\left(1\right)$ arises from pure top-loop mediated $gg\rightarrow H$
amplitudes, the second term proportional to $\tau_{b}^{-1}$ owes to
the interference of top- with bottom-quark loop $gg\rightarrow H$
amplitudes. To evaluate finite mass corrections to the Sudakov form
factor we neglect terms $\mathcal{O}\left(\tau_{b}^{-2}\right)$,
arising from pure $b$-loop mediated contributions.

For the pole $b$-quark-mass, $m_{b}=4.75$ GeV, and $m_{H}=125$ GeV we
find $\frac{3}{4}\,\mathcal{F}_{tb}\,\tau_{b}^{-1}=-0.125$, while
taking instead the $\overline{\mathrm{MS}}$ $b$-quark-mass,
$m_{b}=3.38$ GeV, leads to
$\frac{3}{4}\,\mathcal{F}_{tb}\,\tau_{b}^{-1}=-0.084$.

\subsubsection{$qg\rightarrow Hq$ and $gq\rightarrow Hq$ approximate matrix elements}

Neglecting bottom quark contributions, in the large-$m_{t}$ limit the
$qg\rightarrow Hq$ matrix element is given by,
\begin{equation}
\mathcal{M}_{qg\rightarrow
  Hq}^{\left[\infty,0\right]}=\mathcal{M}_{gg\rightarrow
  H}^{\left[\infty,0\right]}\,\left(\frac{8\pi\alpha_{{\scriptscriptstyle
      \mathrm{S}}}C_{{\scriptscriptstyle
      \mathrm{F}}}}{m_{H}^{4}}\right)\,\frac{t^{2}+s^{2}}{\left|u\right|}\,.\label{eq:M2qgHg-large-mt-no-b}
\end{equation}
Our approximate $2\rightarrow2$ matrix element for $qg\rightarrow Hq$
is obtained by taking the large-$m_{t}$-small-$m_{b}$ limit of the
full $qg\rightarrow Hq$ matrix element including top- and bottom-quark
mediated contributions. We express it in terms of a piece,
$\mathcal{M}_{qg\rightarrow Hq}^{\mathcal{F}}$, which respects
conventional soft and collinear factorisation, and a further remainder
term, $\mathcal{M}_{qg\rightarrow Hq}^{\mathcal{R}}$, which does not:
\begin{equation}
\underset{{\scriptstyle
    m_{b}\rightarrow0}}{\lim_{m_{t}\rightarrow\infty}}\mathcal{M}_{qg\rightarrow
  Hq}^{\left[m_{t},m_{b}\right]}\rightarrow\mathcal{M}_{qg\rightarrow
  Hq}^{\mathcal{A}}=\mathcal{M}_{qg\rightarrow
  Hq}^{\mathcal{F}}+\mathcal{M}_{qg\rightarrow
  Hq}^{\mathcal{R}}\label{eq:M2qgHg-approx-i}
\end{equation}
where, 
\begin{eqnarray}
\mathcal{M}_{qg\rightarrow Hq}^{\mathcal{F}} & = &
\mathcal{M}_{qg\rightarrow
  Hq}^{\left[\infty,0\right]}\,\left(1+\frac{3}{4}\,\tau_{b}^{-1}\,\mathcal{F}_{tb}\right)\,,\label{eq:M2qgHg-approx-ii}\\ \nonumber
\\ \mathcal{M}_{qg\rightarrow Hq}^{\mathcal{R}} & = &
\mathcal{M}_{qg\rightarrow
  Hq}^{\left[\infty,0\right]}\,\frac{m_{H}^{2}}{\left|u\right|+m_{H}^{2}}\,\frac{3}{4}\,\tau_{b}^{-1}\left[-\frac{\left|u\right|}{m_{H}^{2}}\,\mathcal{F}_{tb}+\log^{2}\frac{m_{b}^{2}}{\left|u\right|+\varepsilon}-\frac{4\left|u\right|}{\left|u\right|+m_{H}^{2}}\,\log\,\frac{\left|u\right|}{m_{H}^{2}}\right],\,\qquad\label{eq:M2qgHg-approx-iii}
\end{eqnarray}
 with $\varepsilon=m_{b}^{2}$ inserted by-hand, to regulate the
 potential divergence in the region where our small-$m_{b}$
 approximation breaks down.

We note how, in the approximate matrix element here, the interference
of the top- and bottom-quark loop contribution vanishes at high
$\left|u\right|$.  Since the top-loop amplitude contribution to
$qg\rightarrow Hq$ is manifestly real in the large-$m_{t}$ limit, the
vanishing interference at high $\left|u\right|$ implies the
bottom-quark loop amplitude is pure-imaginary, i.e. the latter is
dominated by real rescattering at high energy.

In the limit of large $u$ (backward-scattering of the incoming quark)
the large-$m_{t}$ limit, severely fails to describe the Higgs boson's
transverse momentum for intermediate $p_{T}$. On the other hand, the
quark-mass dependence of the purely top-quark mediated contribution
can be captured by a relatively simple formula:
\begin{eqnarray}
\mathcal{M}_{qg\rightarrow Hq}^{\left[m_{t},0\right]} & = &
\mathcal{M}_{qg\rightarrow
  Hq}^{\left[\infty,0\right]}\times\frac{36m_{t}^{4}}{\left(m_{H}^{2}+\left|u\right|\right)^{4}}\,\left[-\left|u\right|-m_{H}^{2}\phantom{\left(\frac{1}{2}\sqrt{\frac{\left|u\right|}{m_{t}^{2}}}\right)^{2}}\right.\label{eq:M2qgHg-approx-iv}
  \nonumber\\ \nonumber \\ & &
  -2\left|u\right|\left(\sqrt{\frac{4m_{t}^{2}}{m_{H}^{2}}-1}\sin^{-1}\left(\frac{1}{2}\sqrt{\frac{m_{H}^{2}}{m_{t}^{2}}}\right)-\sqrt{1+\frac{4m_{t}^{2}}{\left|u\right|}}\sinh^{-1}\left(\frac{1}{2}\sqrt{\frac{\left|u\right|}{m_{t}^{2}}}\right)\right)\quad\quad\label{eq:M2qgHg-approx-v}\nonumber\\ &
  &
  \left.+\left(4m_{t}^{2}-s-t\right)\left(\sin^{-1}\left(\frac{1}{2}\sqrt{\frac{m_{H}^{2}}{m_{t}^{2}}}\right)^{2}+\sinh^{-1}\left(\frac{1}{2}\sqrt{\frac{\left|u\right|}{m_{t}^{2}}}\right)^{2}\right)\right]^{2}\,.\label{eq:M2qgHg-approx-vi}
\end{eqnarray}
The approximate matrix element implemented is given by the rescaling
\begin{equation}
\mathcal{M}_{qg\rightarrow
  Hq}^{\mathcal{A}}\rightarrow\frac{\mathcal{M}_{qg\rightarrow
    Hq}^{\left[m_{t},0\right]}}{\mathcal{M}_{qg\rightarrow
    Hq}^{\left[\infty,0\right]}}\times\mathcal{M}_{qg\rightarrow
  Hq}^{\mathcal{A}}\,.
\end{equation}

The analogous $gq\rightarrow Hq$ matrix element is readily obtained by
substituting $u\leftrightarrow t$ in the $qg\rightarrow Hq$ formulae
above.

\subsubsection{$gg\rightarrow Hg$ approximate matrix element}

Neglecting bottom quark contributions, in the large-$m_{t}$ limit the
$gg\rightarrow Hg$ matrix element is given by,
\begin{equation}
\mathcal{M}_{gg\rightarrow
  Hg}^{\left[\infty,0\right]}=\mathcal{M}_{gg\rightarrow
  H}^{\left[\infty,0\right]}\,\left(\frac{8\pi\alpha_{{\scriptscriptstyle
      \mathrm{S}}}\,2N_{{\scriptscriptstyle
      \mathrm{C}}}}{m_{{\scriptscriptstyle
      \mathrm{H}}}^{4}}\right)\,\frac{m_{{\scriptscriptstyle
      \mathrm{H}}}^{8}+s^{4}+t^{4}+u^{4}}{2stu}\,.\label{eq:M2ggHg-large-mt-no-b}
\end{equation}
Omitting terms $\mathcal{O}\left(\tau_{b}^{-2}\right)$, in the
large-$m_{t}$-small-$m_{b}$ approximation, we go on and take the limit
$m_{b}\ll p_{T}\ll m_{H}$, keeping only the resulting leading
$\log^{2}\, m_{b}^{2}$ term in the top-bottom interference piece we
obtain
\begin{equation}
\underset{\underset{{\scriptstyle m_{b}\ll p_{T}\ll
      m_{H}}}{{\scriptstyle
      m_{b}\rightarrow0}}}{\lim_{m_{t}\rightarrow\infty}}\mathcal{M}_{gg\rightarrow
  Hg}^{\left[m_{t},m_{b}\right]}\rightarrow\mathcal{M}_{gg\rightarrow
  Hg}^{\mathcal{F}}+\mathcal{M}_{gg\rightarrow
  Hg}^{\mathcal{R}}\,,\label{eq:M2ggHg-approx-i}
\end{equation}
where,
\begin{eqnarray}
\mathcal{M}_{gg\rightarrow Hg}^{\mathcal{F}} & = &
\mathcal{M}_{gg\rightarrow
  Hg}^{\left[\infty,0\right]}\,\left(1+\frac{3}{4}\,\mathcal{F}_{tb}\,\tau_{b}^{-1}\right)\,,\label{eq:MggHg-approx-i-F}\\ \nonumber
\\ \mathcal{M}_{gg\rightarrow Hg}^{\mathcal{R}} & = &
\mathcal{M}_{gg\rightarrow
  H}^{\left[\infty,0\right]}\,\left(\frac{8\pi\alpha_{{\scriptscriptstyle
      \mathrm{S}}}\,2N_{{\scriptscriptstyle
      \mathrm{C}}}}{p_{T}^{2}}\right)\,\frac{1}{z^{2}}\,\frac{3}{4}\,\tau_{b}^{-1}\,\left(1-z\right)\left(1-z+\frac{5}{4}\,
z^{2}\right)\,\log^{2}\left(\frac{m_{b}^{2}}{p_{T}^{2}}\right)\,.\qquad\label{eq:MggHg-approx-i-R}
\end{eqnarray}

For $z\rightarrow1$ this approximation recovers the results of
conventional soft eikonal factorisation at the level of the pure
$t$-loop mediated contribution, as well as for those terms above
originating from $t$-$b$ interference. The vanishing of the
factorization-breaking remainder pieces as $z\rightarrow1$, is
consistent with Sec.~3.1 of ref.~\cite{Banfi:2013eda} concerning soft
factorisation. The formulae here extend those of
ref.~\cite{Banfi:2013eda}, giving a unified description of soft and
collinear regions.

Starting from the large-$m_{t}$-small-$m_{b}$ approximation and taking
instead just the $z\rightarrow0$ high energy limit, neglecting terms
$\mathcal{O}(1/z)$ gives formulae with a structure strongly resembling
of the large-$m_{t}$-small-$m_{b}$ approximation applied to the
$qg\rightarrow Hq$ and $gq\rightarrow Hq$ channels
eqs.~(\ref{eq:M2qgHg-approx-ii}, \ref{eq:M2qgHg-approx-iii}):
\begin{equation}
\underset{\underset{z\rightarrow0}{{\scriptstyle
      m_{b}\rightarrow0}}}{\lim_{m_{t}\rightarrow\infty}}\mathcal{M}_{gg\rightarrow
  Hg}^{\left[m_{t},m_{b}\right]}\rightarrow\mathcal{M}_{gg\rightarrow
  Hg}^{\prime\mathcal{F}}+\mathcal{M}_{gg\rightarrow
  Hg}^{\prime\mathcal{R}}\,,\label{eq:M2ggHg-approx-ii}
\end{equation}
where,
\begin{eqnarray}
\mathcal{M}_{gg\rightarrow Hg}^{\prime\mathcal{F}} & = &
\mathcal{M}_{gg\rightarrow
  Hg}^{\left[\infty,0\right]}\,\,\left(1+\frac{3}{4}\,\mathcal{F}_{tb}\,\tau_{b}^{-1}\right)\,,\label{eq:M2ggHg-approx-ii-F}\\ \nonumber
\\ \mathcal{M}_{gg\rightarrow Hg}^{\prime\mathcal{R}} & = &
\mathcal{M}_{gg\rightarrow
  Hg}^{\left[\infty,0\right]}\,\frac{m_{H}^{2}}{p_{T}^{2}+m_{H}^{2}}\,\frac{3}{4}\,\tau_{b}^{-1}\,\left[-\frac{p_{T}^{2}}{m_{H}^{2}}\,\mathcal{F}_{tb}+\log^{2}\left(\frac{m_{b}^{2}}{p_{T}^{2}}\right)-\frac{4p_{T}^{2}}{p_{T}^{2}+m_{H}^{2}}\log\left(\frac{p_{T}^{2}}{m_{H}^{2}}\right)\right]\,.\qquad\label{eq:M2ggHg-approx-ii-R}
\end{eqnarray}

Our final approximate matrix element expression for $gg\rightarrow Hg$
is made by combining those for the $m_{b}\ll p_{T}\ll m_{H}$ and
$z\rightarrow0$ domains into a single formula respecting both limits
and interpolating between them:
\begin{equation}
\mathcal{M}_{gg\rightarrow
  Hg}^{\left[m_{t},m_{b}\right]}\rightarrow\mathcal{M}_{gg\rightarrow
  Hg}^{\mathcal{A}}\,=\frac{\mathcal{M}_{gg\rightarrow
    H}^{\left[m_{t},0\right]}}{\mathcal{M}_{gg\rightarrow
    H}^{\left[\infty,0\right]}}\,\left[\mathcal{M}_{gg\rightarrow
    Hg}^{\mathcal{F}}+\mathcal{M}_{gg\rightarrow
    Hg}^{\mathcal{R}}\right]\,,\label{eq:M2ggHg-approx-final}
\end{equation}
with
\begin{eqnarray}
\mathcal{M}_{gg\rightarrow Hg}^{\mathcal{F}} & = &
\mathcal{M}_{gg\rightarrow
  Hg}^{\left[\infty,0\right]}\left(1+\frac{3}{4}\,\mathcal{F}_{tb}\,\tau_{b}^{-1}\right)\,,\label{eq:M2ggHg-approx-final-F}\\ \nonumber
\\ \mathcal{M}_{gg\rightarrow Hg}^{\mathcal{R}} & = &
\mathcal{M}_{gg\rightarrow
  Hg}^{\left[\infty,0\right]}\,\frac{m_{H}^{2}}{p_{T}^{2}+m_{H}^{2}}\,\frac{3}{4}\,\tau_{b}^{-1}\,\label{eq:M2ggHg-approx-final-R}\\ &
\times &
\left[-\frac{p_{T}^{2}}{m_{H}^{2}}\,\mathcal{F}_{tb}+\frac{\frac{1}{z}\left(1-z+\frac{5}{4}\,
    z^{2}\right)}{z\left(1-z\right)+\frac{1-z}{z}+\frac{z}{1-z}}\log^{2}\left(\frac{m_{b}^{2}}{p_{T}^{2}+\varepsilon}\right)-\frac{4p_{{\scriptscriptstyle
        \mathrm{T}}}^{2}}{p_{{\scriptscriptstyle
        \mathrm{T}}}^{2}+m_{{\scriptscriptstyle
        \mathrm{H}}}^{2}}\log\left(\frac{p_{T}^{2}}{m_{H}^{2}}\right)\right]\,.\nonumber
\end{eqnarray}
As in eq.~\eqref{eq:M2qgHg-approx-iii}, $\varepsilon=m_{b}^{2}$ has
been inserted by-hand in eq.~\eqref{eq:M2ggHg-approx-final-R} to
regulate the spurious divergence in the region where our small-$m_{b}$
approximation breaks down. The ratio $\mathcal{M}_{gg\rightarrow
  H}^{\left[m_{t},0\right]}/\mathcal{M}_{gg\rightarrow
  H}^{\left[\infty,0\right]}$ appears in
eq.~\eqref{eq:M2ggHg-approx-final} as an economical bid to capture
some of the finite top-quark mass dependence.


\begin{thebibliography}{10}

\bibitem{Aad:2012tfa}
{\bf ATLAS Collaboration} Collaboration, G.~Aad et~al., {\it {Observation of a
  new particle in the search for the Standard Model Higgs boson with the ATLAS
  detector at the LHC}},  {\em Phys.Lett.} {\bf B716} (2012) 1--29,
  [\href{http://xxx.lanl.gov/abs/1207.7214}{{\tt arXiv:1207.7214}}].

\bibitem{Chatrchyan:2012ufa}
{\bf CMS Collaboration} Collaboration, S.~Chatrchyan et~al., {\it {Observation
  of a new boson at a mass of 125 GeV with the CMS experiment at the LHC}},
  {\em Phys.Lett.} {\bf B716} (2012) 30--61,
  [\href{http://xxx.lanl.gov/abs/1207.7235}{{\tt arXiv:1207.7235}}].

\bibitem{CMS:bxa}
{\bf CMS} Collaboration, {\it {Update on the search for the standard model
  Higgs boson in pp collisions at the LHC decaying to W + W in the fully
  leptonic final state}},  {\em CMS-PAS-HIG-13-003} (2013).

\bibitem{Chatrchyan:2013vaa}
{\bf CMS} Collaboration, S.~Chatrchyan et~al., {\it {Search for a Higgs boson
  decaying into a Z and a photon in pp collisions at $\sqrt{s}$ = 7 and 8
  TeV}},  {\em Phys.Lett.} {\bf B726} (2013) 587--609,
  [\href{http://xxx.lanl.gov/abs/1307.5515}{{\tt arXiv:1307.5515}}].

\bibitem{Chatrchyan:2014nva}
{\bf CMS} Collaboration, S.~Chatrchyan et~al., {\it {Evidence for the 125 GeV
  Higgs boson decaying to a pair of $\tau$ leptons}},  {\em JHEP} {\bf 1405}
  (2014) 104, [\href{http://xxx.lanl.gov/abs/1401.5041}{{\tt
  arXiv:1401.5041}}].

\bibitem{CMS:zwa}
{\bf CMS} Collaboration, {\it {Search for SM Higgs in WH to WWW to 3l 3nu}},
  {\em CMS-PAS-HIG-13-009} (2013).

\bibitem{Khachatryan:2014jba}
{\bf CMS} Collaboration, V.~Khachatryan et~al., {\it {Precise determination of
  the mass of the Higgs boson and tests of compatibility of its couplings with
  the standard model predictions using proton collisions at 7 and 8 TeV}},
  \href{http://xxx.lanl.gov/abs/1412.8662}{{\tt arXiv:1412.8662}}.

\bibitem{ATLAS:2013oma}
{\bf ATLAS} Collaboration, {\it {Measurements of the properties of the
  Higgs-like boson in the two photon decay channel with the ATLAS detector
  using 25 $\mathrm{fb}^{-1}$ of proton-proton collision data}},  {\em
  ATLAS-CONF-2013-012, ATLAS-COM-CONF-2013-015} (2013).

\bibitem{ATLAS:2013qma}
{\bf ATLAS} Collaboration, {\it {Search for a Standard Model Higgs boson in
  $H\rightarrow \mu\mu$ decays with the ATLAS detector.}},  {\em
  ATLAS-CONF-2013-010, ATLAS-COM-CONF-2013-003} (2013).

\bibitem{ATLAS:2013rma}
{\bf ATLAS} Collaboration, {\it {Search for the Standard Model Higgs boson in
  the $H \rightarrow Z\gamma$ decay mode with pp collisions at $\sqrt{s} =$ 7
  and 8 TeV}},  {\em ATLAS-CONF-2013-009, ATLAS-COM-CONF-2013-014} (2013).

\bibitem{ATLAS-CONF-2013-040}
{\it Study of the spin of the new boson with up to 25~fb$^{-1}$ of atlas data},
   Tech. Rep. ATLAS-CONF-2013-040, CERN, Geneva, Apr, 2013.

\bibitem{Aad:2014lwa}
{\bf ATLAS Collaboration} Collaboration, G.~Aad et~al., {\it {Measurements of
  fiducial and differential cross sections for Higgs boson production in the
  diphoton decay channel at $\sqrt{s}=8$ TeV with ATLAS}},  {\em JHEP} {\bf
  1409} (2014) 112, [\href{http://xxx.lanl.gov/abs/1407.4222}{{\tt
  arXiv:1407.4222}}].

\bibitem{Aad:2014tca}
{\bf ATLAS Collaboration} Collaboration, G.~Aad et~al., {\it {Fiducial and
  differential cross sections of Higgs boson production measured in the
  four-lepton decay channel in $pp$ collisions at $\sqrt{s}$=8 TeV with the
  ATLAS detector}},  {\em Phys.Lett.} {\bf B738} (2014) 234--253,
  [\href{http://xxx.lanl.gov/abs/1408.3226}{{\tt arXiv:1408.3226}}].

\bibitem{Frixione:2002ik}
S.~Frixione and B.~R. Webber, {\it {Matching NLO QCD computations and parton
  shower simulations}},  {\em JHEP} {\bf 0206} (2002) 029,
  [\href{http://xxx.lanl.gov/abs/hep-ph/0204244}{{\tt hep-ph/0204244}}].

\bibitem{Nason:2004rx}
P.~Nason, {\it {A New method for combining NLO QCD with shower Monte Carlo
  algorithms}},  {\em JHEP} {\bf 0411} (2004) 040,
  [\href{http://xxx.lanl.gov/abs/hep-ph/0409146}{{\tt hep-ph/0409146}}].

\bibitem{Frixione:2007vw}
S.~Frixione, P.~Nason, and C.~Oleari, {\it {Matching NLO QCD computations with
  Parton Shower simulations: the POWHEG method}},  {\em JHEP} {\bf 0711} (2007)
  070, [\href{http://xxx.lanl.gov/abs/0709.2092}{{\tt arXiv:0709.2092}}].

\bibitem{Hamilton:2013fea}
K.~Hamilton, P.~Nason, E.~Re, and G.~Zanderighi, {\it {NNLOPS simulation of
  Higgs boson production}},  {\em JHEP} {\bf 1310} (2013) 222,
  [\href{http://xxx.lanl.gov/abs/1309.0017}{{\tt arXiv:1309.0017}}].

\bibitem{Hoche:2014dla}
S.~Hoche, Y.~Li, and S.~Prestel, {\it {Higgs-boson production through gluon
  fusion at NNLO QCD with parton showers}},  {\em Phys.Rev.} {\bf D90} (2014),
  no.~5 054011, [\href{http://xxx.lanl.gov/abs/1407.3773}{{\tt
  arXiv:1407.3773}}].

\bibitem{Hoeche:2014aia}
S.~Hoeche, Y.~Li, and S.~Prestel, {\it {Drell-Yan lepton pair production at
  NNLO QCD with parton showers}},
  \href{http://xxx.lanl.gov/abs/1405.3607}{{\tt arXiv:1405.3607}}.

\bibitem{Karlberg:2014qua}
A.~Karlberg, E.~Re, and G.~Zanderighi, {\it {NNLOPS accurate Drell-Yan
  production}},  {\em JHEP} {\bf 1409} (2014) 134,
  [\href{http://xxx.lanl.gov/abs/1407.2940}{{\tt arXiv:1407.2940}}].

\bibitem{Hamilton:2012rf}
K.~Hamilton, P.~Nason, C.~Oleari, and G.~Zanderighi, {\it {Merging H/W/Z + 0
  and 1 jet at NLO with no merging scale: a path to parton shower + NNLO
  matching}},  {\em JHEP} {\bf 1305} (2013) 082,
  [\href{http://xxx.lanl.gov/abs/1212.4504}{{\tt arXiv:1212.4504}}].

\bibitem{Graudenz:1992pv}
D.~Graudenz, M.~Spira, and P.~Zerwas, {\it {QCD corrections to Higgs boson
  production at proton proton colliders}},  {\em Phys.Rev.Lett.} {\bf 70}
  (1993) 1372--1375.

\bibitem{Spira:1995rr}
M.~Spira, A.~Djouadi, D.~Graudenz, and P.~Zerwas, {\it {Higgs boson production
  at the LHC}},  {\em Nucl.Phys.} {\bf B453} (1995) 17--82,
  [\href{http://xxx.lanl.gov/abs/hep-ph/9504378}{{\tt hep-ph/9504378}}].

\bibitem{Harlander:2005rq}
R.~Harlander and P.~Kant, {\it {Higgs production and decay: Analytic results at
  next-to-leading order QCD}},  {\em JHEP} {\bf 0512} (2005) 015,
  [\href{http://xxx.lanl.gov/abs/hep-ph/0509189}{{\tt hep-ph/0509189}}].

\bibitem{Anastasiou:2006hc}
C.~Anastasiou, S.~Beerli, S.~Bucherer, A.~Daleo, and Z.~Kunszt, {\it {Two-loop
  amplitudes and master integrals for the production of a Higgs boson via a
  massive quark and a scalar-quark loop}},  {\em JHEP} {\bf 0701} (2007) 082,
  [\href{http://xxx.lanl.gov/abs/hep-ph/0611236}{{\tt hep-ph/0611236}}].

\bibitem{Aglietti:2006tp}
U.~Aglietti, R.~Bonciani, G.~Degrassi, and A.~Vicini, {\it {Analytic Results
  for Virtual QCD Corrections to Higgs Production and Decay}},  {\em JHEP} {\bf
  0701} (2007) 021, [\href{http://xxx.lanl.gov/abs/hep-ph/0611266}{{\tt
  hep-ph/0611266}}].

\bibitem{Bonciani:2007ex}
R.~Bonciani, G.~Degrassi, and A.~Vicini, {\it {Scalar particle contribution to
  Higgs production via gluon fusion at NLO}},  {\em JHEP} {\bf 0711} (2007)
  095, [\href{http://xxx.lanl.gov/abs/0709.4227}{{\tt arXiv:0709.4227}}].

\bibitem{Ellis:1987xu}
R.~K. Ellis, I.~Hinchliffe, M.~Soldate, and J.~van~der Bij, {\it {Higgs Decay
  to tau+ tau-: A Possible Signature of Intermediate Mass Higgs Bosons at the
  SSC}},  {\em Nucl.Phys.} {\bf B297} (1988) 221.

\bibitem{Baur:1989cm}
U.~Baur and E.~N. Glover, {\it {Higgs Boson Production at Large Transverse
  Momentum in Hadronic Collisions}},  {\em Nucl.Phys.} {\bf B339} (1990)
  38--66.

\bibitem{Harlander:2012hf}
R.~V. Harlander, T.~Neumann, K.~J. Ozeren, and M.~Wiesemann, {\it {Top-mass
  effects in differential Higgs production through gluon fusion at order
  $\alpha_s^4$}},  {\em JHEP} {\bf 1208} (2012) 139,
  [\href{http://xxx.lanl.gov/abs/1206.0157}{{\tt arXiv:1206.0157}}].

\bibitem{MCFM}
{\it {MCFM -- Monte Carlo for FeMtobarn processes: http://mcfm.fnal.gov}}, .

\bibitem{Anastasiou:2005qj}
C.~Anastasiou, K.~Melnikov, and F.~Petriello, {\it {Fully differential Higgs
  boson production and the di-photon signal through next-to-next-to-leading
  order}},  {\em Nucl.Phys.} {\bf B724} (2005) 197--246,
  [\href{http://xxx.lanl.gov/abs/hep-ph/0501130}{{\tt hep-ph/0501130}}].

\bibitem{Anastasiou:2009kn}
C.~Anastasiou, S.~Bucherer, and Z.~Kunszt, {\it {HPro: A NLO Monte-Carlo for
  Higgs production via gluon fusion with finite heavy quark masses}},  {\em
  JHEP} {\bf 0910} (2009) 068, [\href{http://xxx.lanl.gov/abs/0907.2362}{{\tt
  arXiv:0907.2362}}].

\bibitem{Harlander:2012pb}
R.~V. Harlander, S.~Liebler, and H.~Mantler, {\it {SusHi: A program for the
  calculation of Higgs production in gluon fusion and bottom-quark annihilation
  in the Standard Model and the MSSM}},  {\em Computer Physics Communications}
  {\bf 184} (2013) 1605--1617, [\href{http://xxx.lanl.gov/abs/1212.3249}{{\tt
  arXiv:1212.3249}}].

\bibitem{Bagnaschi:2011tu}
E.~Bagnaschi, G.~Degrassi, P.~Slavich, and A.~Vicini, {\it {Higgs production
  via gluon fusion in the POWHEG approach in the SM and in the MSSM}},  {\em
  JHEP} {\bf 1202} (2012) 088, [\href{http://xxx.lanl.gov/abs/1111.2854}{{\tt
  arXiv:1111.2854}}].

\bibitem{Buschmann:2014sia}
M.~Buschmann, D.~Goncalves, S.~Kuttimalai, M.~Schonherr, F.~Krauss, et~al.,
  {\it {Mass Effects in the Higgs-Gluon Coupling: Boosted vs Off-Shell
  Production}},  \href{http://xxx.lanl.gov/abs/1410.5806}{{\tt
  arXiv:1410.5806}}.

\bibitem{Mantler:2012bj}
H.~Mantler and M.~Wiesemann, {\it {Top- and bottom-mass effects in hadronic
  Higgs production at small transverse momenta through LO+NLL}},  {\em
  Eur.Phys.J.} {\bf C73} (2013), no.~6 2467,
  [\href{http://xxx.lanl.gov/abs/1210.8263}{{\tt arXiv:1210.8263}}].

\bibitem{Harlander:2014uea}
R.~V. Harlander, H.~Mantler, and M.~Wiesemann, {\it {Transverse momentum
  resummation for Higgs production via gluon fusion in the MSSM}},  {\em JHEP}
  {\bf 1411} (2014) 116, [\href{http://xxx.lanl.gov/abs/1409.0531}{{\tt
  arXiv:1409.0531}}].

\bibitem{Grazzini:2013mca}
M.~Grazzini and H.~Sargsyan, {\it {Heavy-quark mass effects in Higgs boson
  production at the LHC}},  {\em JHEP} {\bf 1309} (2013) 129,
  [\href{http://xxx.lanl.gov/abs/1306.4581}{{\tt arXiv:1306.4581}}].

\bibitem{Banfi:2013eda}
A.~Banfi, P.~F. Monni, and G.~Zanderighi, {\it {Quark masses in Higgs
  production with a jet veto}},  {\em JHEP} {\bf 1401} (2014) 097,
  [\href{http://xxx.lanl.gov/abs/1308.4634}{{\tt arXiv:1308.4634}}].

\bibitem{Campbell:2012am}
J.~M. Campbell, R.~K. Ellis, R.~Frederix, P.~Nason, C.~Oleari, et~al., {\it
  {NLO Higgs Boson Production Plus One and Two Jets Using the POWHEG BOX,
  MadGraph4 and MCFM}},  {\em JHEP} {\bf 1207} (2012) 092,
  [\href{http://xxx.lanl.gov/abs/1202.5475}{{\tt arXiv:1202.5475}}].

\bibitem{Hamilton:2012np}
K.~Hamilton, P.~Nason, and G.~Zanderighi, {\it {MINLO: Multi-Scale Improved
  NLO}},  {\em JHEP} {\bf 1210} (2012) 155,
  [\href{http://xxx.lanl.gov/abs/1206.3572}{{\tt arXiv:1206.3572}}].

\bibitem{Catani:2001cc}
S.~Catani, F.~Krauss, R.~Kuhn, and B.~Webber, {\it {QCD matrix elements +
  parton showers}},  {\em JHEP} {\bf 0111} (2001) 063,
  [\href{http://xxx.lanl.gov/abs/hep-ph/0109231}{{\tt hep-ph/0109231}}].

\bibitem{Altarelli:1977zs}
G.~Altarelli and G.~Parisi, {\it {Asymptotic Freedom in Parton Language}},
  {\em Nucl.Phys.} {\bf B126} (1977) 298.

\bibitem{Martin:2009iq}
A.~Martin, W.~Stirling, R.~Thorne, and G.~Watt, {\it {Parton distributions for
  the LHC}},  {\em Eur.Phys.J.} {\bf C63} (2009) 189--285,
  [\href{http://xxx.lanl.gov/abs/0901.0002}{{\tt arXiv:0901.0002}}].

\bibitem{Banfi:2012jm}
A.~Banfi, P.~F. Monni, G.~P. Salam, and G.~Zanderighi, {\it {Higgs and Z-boson
  production with a jet veto}},  {\em Phys.Rev.Lett.} {\bf 109} (2012) 202001,
  [\href{http://xxx.lanl.gov/abs/1206.4998}{{\tt arXiv:1206.4998}}].

\end{thebibliography}

\providecommand{\href}[2]{#2}\begingroup\raggedright\endgroup

\end{document}